\RequirePackage{amsmath}

\documentclass[epjST]{svjour}

\usepackage{amssymb,cite,fullpage,graphicx}

\usepackage[greek,english]{babel}

\usepackage[T1]{fontenc}

\usepackage[utf8]{inputenc}

\usepackage[pdftex]{hyperref}
\hypersetup{
bookmarksnumbered=true,
colorlinks=true,
pdfauthor={\textcopyright~Fabio Clementi},
pdfencoding=unicode,
pdfkeywords={Generalized statistical mechanics, maximum entropy principle, κ-generalized models, distribution of income and wealth},
pdfstartview=FitH,
pdftitle={κ-generalized models of income and wealth distributions: A survey},
unicode=true
}

\usepackage[bf,hang]{subfigure}

\usepackage[nottoc]{tocbibind}

\DeclareMathOperator{\arcsinh}{arcsinh}

\def\makeheadbox{\relax}

\begin{document}

% FRONT MATTER -------------------------------------------------------------------------------------------

\title{$\boldsymbol{\kappa}$-generalized models of income and wealth distributions}

\subtitle{A survey}

\author{F. Clementi\inst{1}\fnmsep\thanks{\textit{Corresponding author}: \href{mailto:fabio.clementi@unimc.it}{\texttt{fabio.clementi@unimc.it}}} \and M. Gallegati\inst{2} \and G. Kaniadakis\inst{3} \and S. Landini\inst{4}\fnmsep\thanks{Views and opinions are those of the author and do not involve the responsibility of IRES Piemonte.}}

\institute{Dipartimento di Scienze Politiche, della Comunicazione e delle Relazioni Internazionali, Universit\`{a} degli Studi di Macerata,  Piazza San Vincenzo Maria Strambi 1, I--62100 Macerata, Italy \and Dipartimento di Scienze Economiche e Sociali, Universit\`{a} Politecnica delle Marche, Piazzale Raffaele Martelli 8, I--60121 Ancona, Italy \and Dipartimento Scienza Applicata e Tecnologia, Politecnico di Torino, Corso Duca degli Abruzzi 24, I--10129 Torino, Italy \and IRES Piemonte, Istituto di Ricerche Economico Sociali del Piemonte, Via Nizza 18, I--10125 Torino, Italy}

\abstract{The paper provides a survey of results related to the ``$\kappa$-generalized distribution'', a statistical model for the size distribution of income and wealth. Topics include, among others, discussion of basic analytical properties, interrelations with other statistical distributions as well as aspects that are of special interest in the income distribution field, such as the Gini index and the Lorenz curve. An extension of the basic model that is most able to accommodate the special features of wealth data is also reviewed. The survey of empirical applications given in this paper shows the $\kappa$-generalized models of income and wealth to be in excellent agreement with the observed data in many cases.}

\maketitle

% INTRODUCTION -------------------------------------------------------------------------------------------

\section{Introduction}
\label{sec:Introduction}

A renewed interest in studying the distribution of income and wealth has emerged over the last decade in both the physics \cite{ChatterjeeYarlagaddaChakrabarti2005,Yakovenko2009,YakovenkoBarkleyRosser2009,ChakrabartiChakrabortiChakravartyChatterjee2013} and economics \cite{Stiglitz2012,Piketty2014,Atkinson2015,Stiglitz2015} communities. The focus has been mostly put on empirical analysis of extensive data sets to infer the exact shape of income and wealth distributions and to design theoretical models that can reproduce them.

A natural starting point in this area of inquiry was the observation that the number of persons in a population whose incomes exceed $x$ is often well approximated by $Cx^{-\alpha}$, for some real $C$ and positive $\alpha$, as Pareto \cite{Pareto1895,Pareto1896,Pareto1897a,Pareto1897b} argued over 100 years ago. Since the early studies of Pareto, numerous empirical works have shown that the power-law tail is an ubiquitous feature of income and wealth distributions. However, even 100 years after Pareto observation, the understanding of the shape of income/wealth distribution is still far to be complete and definitive. This reflects the fact that there are two distributions, one for the rich, following the Pareto power law, and one for the vast majority of people, which appears to be governed by a completely different law.

Over the years, research in the field has considered a wide variety of functional forms as possible models for the size distribution of income and wealth \cite{KleiberKotz2003}, some of which aim at providing a unified framework for the description of real-world data\textemdash including the heavy tails present in empirical income and wealth distributions. Among these, the ``$\kappa$-generalized distribution'' \cite{ClementiGallegatiKaniadakis2007,ClementiDiMatteoGallegatiKaniadakis2008,ClementiGallegatiKaniadakis2009,ClementiGallegatiKaniadakis2010,ClementiGallegatiKaniadakis2012a,ClementiGallegatiKaniadakis2012b} was found to work remarkably well. First proposed in 2007, and further developed over successive years, this model finds its roots in the framework of $\kappa$-generalized statistical mechanics \cite{Kaniadakis2001,Kaniadakis2002,Kaniadakis2005,Kaniadakis2009a,Kaniadakis2009b,Kaniadakis2013} and has a bulk very close to the Weibull distribution, while its upper tail decays following a Pareto power law for high values of income and wealth, hence providing a kind of compromise between the two descriptions.

Given the renewed focus on distributional issues, and motivated by the feeling that fruitful interaction between the two fields of statistical mechanics and economics may suggest a new path for investigating economic relations, it is the aim of the present paper to present a survey of parametric models belonging to the ``$\kappa$-generalized family'', a set of fitting functions for the size distribution of income and wealth that were typically found as superior alternatives to other widely used parametric models. It will be given knowledge of their basic statistical properties and evidence of their empirical plausibility, hoping to convince readers that this class of distributions deserves attention and study.

The paper is organized as follows. In view of its importance for the proposed statistical models, Section \ref{sec:TheKappaExponentialFunctionAndItsRelativisticOrigin} reviews the relativistic mechanism originating the $\kappa$-exponential function and discusses some of its basic properties. A survey of the $\kappa$-generalized models of income and wealth distribution, including interrelations with other distributions as well as basic statistical properties and inferential aspects, is provided in Section \ref{sec:UseOfTheKappaGeneralizedStatisticsForModelingIncomeAndWealthDistributions}. In Section \ref{sec:ApplicationsOfKappaGeneralizedModelsToIncomeAndWealthData} we present a survey of empirical applications that show how $\kappa$-generalized models of income and wealth are in excellent agreement with the observed data in many cases. Section \ref{sec:DeformToGoStrangeAnExplanationOfWhyKappaGeneralizedModelsAreAGoodFitToIncomeAndWealthDistributions} aims at providing an explanation about why $\kappa$-generalized models provide a better fit to empirical
distributions than other widely used parametric models for the distribution of income and wealth. Finally, in Section \ref{sec:ConcludingRemarks} some concluding remarks are reported.

% THE κ-EXPONENTIAL FUNCTION AND ITS RELATIVISTIC ORIGIN -------------------------------------------------

\section{\texorpdfstring{The $\boldsymbol{\kappa}$-exponential function and its relativistic origin}{The κ-exponential function and its relativistic origin}}
\label{sec:TheKappaExponentialFunctionAndItsRelativisticOrigin}

In experimental particle physics it is well known that the particle distribution at low energies is described by the Boltzmann-Maxwell exponential distribution, i.e.
\begin{equation}
f\left(E\right)\underset{E\rightarrow0}{\sim}\exp\left(-\beta E\right).
\label{eq:Equation_A1}
\end{equation}
On the contrary, at high energies the power-law tailed distribution
\begin{equation}
f\left(E\right)\underset{E\rightarrow+\infty}{\sim}E^{-1/\kappa}
\label{eq:Equation_A2}
\end{equation}
has been observed systematically in the last six decades. This dichotomy in the behavior of the distribution function at low and high energies has stimulated the theoretical research to find a distribution function able to describe the experimental observations on the entire range of energies. Clearly this distribution function had to be motivated by solid and indisputable first principles. In the early 2000s a new proposal was advanced which immediately attracted the attention of many researchers \cite{Kaniadakis2002,Kaniadakis2005,Kaniadakis2009a,Kaniadakis2009b,Trivellato2009,ImparatoTrivellato2010,Trivellato2012,Kaniadakis2013,Trivellato2013,MorettoPasqualiTrivellato2016}. 
The new proposed distribution is given by
\begin{equation}
f\left(E\right)=N\exp_{\kappa}\left(-\beta E+\beta\mu\right),
\label{eq:Equation_A3}
\end{equation}
where $\beta=1/K_{B}T$ is proportional to the reciprocal of the system temperature, $\mu$ is the chemical potential, and the $\kappa$-exponential function $\exp_{\kappa}\left(x\right)$ is defined as
\begin{equation}
\exp_{\kappa}\left(x\right)=\left(\sqrt{1+\kappa^{2}x^{2}}+\kappa x\right)^{1/\kappa},\quad 0<\kappa^{2}<1.
\label{eq:Equation_A4}
\end{equation}
The new distribution at low energies behaves like Boltzmann-Maxwell distribution \eqref{eq:Equation_A1} while at high energies present the power-law tails \eqref{eq:Equation_A2}. The more interesting feature of the new distribution \eqref{eq:Equation_A3} is that the roots of $\kappa$-exponential function $\exp_{\kappa}\left(x\right)$ given in \eqref{eq:Equation_A4} can be found easily within the Einstein special relativity. In other words, the the function $\exp_{\kappa}\left(x\right)$ emerges as a relativistic deformation of the Euler ordinary exponential $\exp\left(x\right)$ of the linear classical physics. The deformation parameter $\kappa$ is proportional to the reciprocal of light speed $c$, so that the mechanism generating the $\exp_{\kappa}\left(x\right)$ is simply the one established by the Lorentz transformations of special relativity.

In the following we briefly recall the relativistic mechanism originating the $\kappa$-exponential function and discuss its main mathematical properties \cite{Kaniadakis2002,Kaniadakis2005,Kaniadakis2009a,Kaniadakis2009b,Kaniadakis2013}. Let us consider in the one-dimension inertial frame $\mathcal{S}$ two identical particles $A$ and $B$, of rest mass $m$. We suppose that the two particles move, in opposite directions, with velocities $v_{A}$ and $v_{B}$ respectively. The momenta of the two particles are indicated with $p_{A}=p\left(v{_A}\right)$ and $p_{B}=p\left(v_{B}\right)$ respectively, where
\begin{equation}
p\left(v\right)=\frac{mv}{\sqrt{1-v^{2}/c^{2}}}.
\label{eq:Equation_A5}
\end{equation}
We consider now the same particles in the rest frame $\mathcal{S}'$ of particle $B$. In this new frame the velocity and momentum of the particle $B$ are $v'_{B}=0$ and $p'_{B}=0$ respectively. In $\mathcal{S}'$ the velocity of particle $A$ is given by the Einstein law
\begin{equation}
v'_{A}=\frac{v_{A}+v_{B}}{1+v_{A}v_{B}/c^{2}},
\label{eq:Equation_A6}
\end{equation}
defining the relativistic velocity composition law, which follows directly from the kinematic Lorentz transformations.

In order to obtain the additivity law for the relativistic momenta, we first invert \eqref{eq:Equation_A5} so as to have
\begin{equation}
v=\frac{p/m}{\sqrt{1+p^{2}/m^{2}c^{2}}}.
\label{eq:Equation_A7}
\end{equation}
Then, after substitution of the velocity \eqref{eq:Equation_A7} in the velocity additivity law expressed by \eqref{eq:Equation_A6}, we obtain the following additivity law for the relativistic momenta
\begin{equation}
p'_{A}=p_{A}\sqrt{1+p_{B}^{2}/m^{2}c^{2}}+p_{B}\sqrt{1+p_{A}^{2}/m^{2}c^{2}}.
\label{eq:Equation_A8}
\end{equation}

Let us now introduce in place of the dimensional variables $\left(v,p\right)$ the dimensionless variables $\left(u,q\right)$ through $v/u=p/mq=\left|\kappa\right|c=v_{\ast}<c$, with $0<\kappa^{2}<1$. It is easy to verify that the additivity law \eqref{eq:Equation_A8} in terms of the dimensionless momenta $q$ assumes the form
\begin{equation}
q'_{A}=q_{A}\sqrt{1+\kappa^{2}q_{B}^{2}}+q_{B}\sqrt{1+\kappa^{2}q_{A}^{2}}.
\end{equation}
The latter transformation law can be written in the following factorized form
\begin{equation}
\exp_{\kappa}\left(q'_{A}\right)=\exp_{\kappa}\left(q_{A}\right)\exp_{\kappa}\left(q_{B}\right),
\label{eq:Equation_A10}
\end{equation}
where the function $\exp_{\kappa}\left(\cdot\right)$ is the $\kappa$-exponential function given by \eqref{eq:Equation_A4}. It follows that the $\kappa$-exponential can be viewed as the function permitting to write the additivity law \eqref{eq:Equation_A8} of relativistic momenta in the factorized form \eqref{eq:Equation_A10}.

From the definition of $\exp_{\kappa}\left(x\right)$ it results that
\begin{align}
&\exp_{0}\left(x\right)\equiv\lim_{\kappa\rightarrow0}\exp_{\kappa}\left(x\right)=\exp\left(x\right),\\
&\exp_{-\kappa}\left(x\right)=\exp_{\kappa}\left(x\right).
\end{align}
Like the ordinary exponential, $\exp_{\kappa}\left(x\right)$ has the properties
\begin{align}
&\exp_{\kappa}\left(x\right)\in C^{\infty}\left(\mathbf{R}\right),\\
&\frac{\operatorname{d}}{\operatorname{d}x}\exp_{\kappa}\left(x\right)>0,\\
&\exp_{\kappa}\left(-\infty\right)=0^{+},\\
&\exp_{\kappa}\left(0\right)=1,\\
&\exp_{\kappa}\left(+\infty\right)=+\infty,\\
&\exp_{\kappa}\left(x\right)\exp_{\kappa}\left(-x\right)=1.
\label{eq:Equation_A18}
\end{align}

The property \eqref{eq:Equation_A18} emerges as particular case of the more general one
\begin{equation}
\exp_{\kappa}\left(x\right)\exp_{\kappa}\left(y\right)=\exp_{\kappa}\left(x\stackrel{\kappa}{\oplus}y\right),
\end{equation}
where
\begin{equation}
x\stackrel{\kappa}{\oplus}y=x\sqrt{1+\kappa^{2}y^{2}}+y\sqrt{1+\kappa^{2}x^{2}},
\end{equation}

Furthermore, $\exp_{\kappa}\left(x\right)$ has the property
\begin{equation}
\left[\exp_{\kappa}\left(x\right)\right]^{r}=\exp_{\kappa/r}\left(rx\right),
\end{equation}
with $r\in\mathbf{R}$, which in the limit $\kappa\rightarrow0$ reproduces one well known property of the ordinary exponential.

We remark the following convexity property
\begin{equation}
\frac{\operatorname{d}^2}{\operatorname{d}x^{2}}\exp_{\kappa}\left(x\right)>0,\quad x\in\mathbf{R},
\end{equation}
holding when $\kappa^{2}<1$.

Undoubtedly, one of the more interesting properties of $\exp_{\kappa}\left(x\right)$ is its power-law asymptotic behavior
\begin{equation}
\exp_{\kappa}\left(x\right)\underset{x\rightarrow\pm\infty}{\sim}\left|2\kappa x\right|^{\pm1/\left|\kappa\right|}.
\end{equation}

The Taylor expansion of $\exp_{\kappa}\left(x\right)$ can be written in the following form
\begin{equation}
\exp_{\kappa}\left(x\right)=\sum_{n=0}^{\infty}\xi_{n}\left(\kappa\right)\frac{x^{n}}{n!},\quad\kappa^{2}x^{2}<1,
\end{equation}
where the polynomials $\xi_{n}\left(\kappa\right)$ are given by
\begin{align}
&\xi_{2m}\left(\kappa\right)=\prod_{j=0}^{m-1}\left[1-\left(2j\right)^{2}\kappa^{2}\right],\quad m>0,\\
&\xi_{2m+1}\left(\kappa\right)=\prod_{j=0}^{m-1}\left[1-\left(2j+1\right)^{2}\kappa^{2}\right],\quad m>0,
\end{align}
and can be generated by the following simple recursive formula
\begin{align}
&\xi_{n+2}\left(\kappa\right)=\left(1-n^{2}\kappa^{2}\right)\xi_{n}\left(\kappa\right),\quad n\geq0,\\
&\xi_{0}\left(\kappa\right)=\xi_{1}\left(\kappa\right)=1.
\end{align}
It is remarkable that the first three terms in the Taylor expansion of $\exp_{\kappa}\left(x\right)$ are the same of the ordinary exponential
\begin{equation}
\begin{split}
\exp_{\kappa}\left(x\right)=&1+x+\frac{x^{2}}{2}+\left(1-\kappa^{2}\right)\frac{x^{3}}{3!}+\left(1-4\kappa^{2}\right)\frac{x^{4}}{4!}\\
&+\left(1-\kappa^{2}\right)\left(1-9\kappa^{2}\right)\frac{x^{5}}{5!}+\left(1-4\kappa^{2}\right)\left(1-16\kappa^{2}\right)\frac{x^{6}}{6!}+\cdots.
\end{split}
\end{equation}

% USE OF THE κ-GENERALIZED STATISTICS FOR MODELING INCOME AND WEALTH DISTRIBUTIONS -----------------------

\section{\texorpdfstring{Use of the $\boldsymbol{\kappa}$-generalized statistics for modeling income and wealth distributions}{Use of the κ-generalized statistics for modeling income and wealth distributions}}
\label{sec:UseOfTheKappaGeneralizedStatisticsForModelingIncomeAndWealthDistributions}

% The κ-generalized model for income distributions -------------------------------------------------------

\subsection{The $\boldsymbol{\kappa}$-generalized model for income distributions}
\label{sec:TheKappaGeneralizedModelForIncomeDistributions}

% Definitions

\subsubsection{Definitions}
\label{sec:Definitions}

In recent years the $\kappa$-deformed exponential \eqref{eq:Equation_A4} was adopted successfully to analyze also non-physical systems, including economic systems. In particular, the $\kappa$-deformation has been employed to study differentiated product markets \cite{RajaoarisonBolducJayet2006,Rajaoarison2008}, in finance \cite{Trivellato2009,ImparatoTrivellato2010,Trivellato2012,Trivellato2013,MorettoPasqualiTrivellato2016}, and for modeling the size distribution of personal income \cite{ClementiGallegatiKaniadakis2007,ClementiDiMatteoGallegatiKaniadakis2008,ClementiGallegatiKaniadakis2009,ClementiGallegatiKaniadakis2010,ClementiGallegatiKaniadakis2012a}. In the latter application a new parametric model, named \textit{$\kappa$-generalized distribution} after \cite{ClementiGallegatiKaniadakis2007}, was defined in terms of the following cumulative distribution function (CDF)
\begin{equation}
F\left(x;\alpha,\beta,\kappa\right)=1-\exp_{\kappa}\left[-\left(\frac{x}{\beta}\right)^{\alpha}\right],\quad x>0,\quad \alpha,\beta>0,\quad \kappa\in\left[0,1\right),
\label{eq:Equation_1}
\end{equation}
where $x$ denotes the income variable and $\left\{\alpha,\beta,\kappa\right\}$ are parameters. The corresponding probability density function (PDF) reads as
\begin{equation}
f\left(x;\alpha,\beta,\kappa\right)=\frac{\alpha}{\beta}\left(\frac{x}{\beta}\right)^{\alpha-1}\frac{\exp_{\kappa}\left[-\left(\frac{x}{\beta}\right)^{\alpha}\right]}{\sqrt{1+\kappa^{2}\left(\frac{x}{\beta}\right)^{2\alpha}}}.
\label{eq:Equation_2}
\end{equation}
References \cite{ClementiGallegatiKaniadakis2007,ClementiDiMatteoGallegatiKaniadakis2008,ClementiGallegatiKaniadakis2009} give slightly different expressions for the $\kappa$-generalized CDF and PDF; their expressions are simply re-parametrized versions of \eqref{eq:Equation_1} and \eqref{eq:Equation_2}.

% Genesis

\subsubsection{Genesis}
\label{sec:Genesis}

As shown by \cite{Landini2016}, the $\kappa$-generalized model of income distribution naturally emerges within the field of $\kappa$-deformed analysis.

Let $X:\Omega_n\rightarrow\mathbb{X}=[0<x_{\min},x_{\max}\ll+\infty]$ be a random variable denoting income. Let also $G\left(x\right)\in\mathcal{C}^{\infty}\left(\mathbb{R}\right)$ be a strictly increasing, invertible and odd function so as to have $G\left(x\right)\approx x$ for $x\rightarrow x_{\min}$, and let $\kappa\in\left[0,1\right)$ be a deformation parameter such that $g_{\kappa}\left(x\right)=G\left(\kappa x\right)$ is a $\kappa$-deformed generator. The \textit{generative representation} of the $\kappa$-deformed random variable $X_{\kappa}$ is a function from $\mathbb{X}$ into its deformation $\mathbb{X}_{\kappa}$, namely
\begin{equation}
X_{\kappa}\equiv\frac{1}{\kappa}\arcsinh\left[g_{\kappa}\left(X\right)\right]=\frac{1}{\kappa}\ln\left[\sqrt{1+g_\kappa^{2}\left(X\right)}+g_{\kappa}\left(X\right)\right].
\label{eq:Equation_3}
\end{equation} 
Notice that the above is not a proper transformation but rather a deformation, as it aims at deforming the quantity $X$ to be regularly treated according to $\kappa$-deformed operators. Among these, the generalized $\kappa$-deformed exponential follows readily from \eqref{eq:Equation_3} as \cite{Kaniadakis2001,Kaniadakis2002,Kaniadakis2005,Kaniadakis2013}
\begin{equation}
\exp_{\kappa}\left(x\right)\equiv\left[\sqrt{1+g_{\kappa}^2\left(x\right)}+g_{\kappa}\left(x\right)\right]^{\frac{1}{\kappa}}=\exp\left\{\frac{1}{\kappa}\arcsinh\left[g_{\kappa}\left(x\right)\right]\right\}=\exp\left(x_{\kappa}\right).
\label{eq:Equation_4}
\end{equation}

Let now $\alpha,\beta>0$ be two additional parameters and define a \textit{regular transformation} as
\begin{equation}
y=\left(\frac{x}{\beta}\right)^\alpha >0\quad\Longrightarrow\quad\operatorname{d}y=\frac{\alpha}{\beta}\left(\frac{x}{\beta}\right)^{\alpha-1}\operatorname{d}x.
\label{eq:Equation_5}
\end{equation}
Due to self-duality of the $\kappa$-deformed exponential, multiplying by $-1$ and applying the operator \eqref{eq:Equation_4} one gets
\begin{equation}
\exp_{\kappa}\left(-y\right)=\sqrt{1+\kappa^{2}y^{2}}-\kappa y,
\end{equation}
which defines both the $\kappa$-deformed distribution,
\begin{equation}
U_{\kappa}\left(y\right)=1-\exp_{\kappa}\left(-y\right),
\end{equation}
and its density,
\begin{equation}
u_{\kappa}\left(y\right)=\frac{\exp_{\kappa}\left(-y\right)}{\sqrt{1+\kappa^{2}y^{2}}}.
\label{eq:Equation_8}
\end{equation}
Therefore, substituting \eqref{eq:Equation_5} into \eqref{eq:Equation_8} gives
\begin{equation}
u_{\kappa}\left[\left(\frac{x}{\beta}\right)^\alpha\right]=\frac{\operatorname{d}U_{\kappa}\left[\left(\frac{x}{\beta}\right)^{\alpha}\right]}{\frac{\alpha}{\beta}\left(\frac{x}{\beta}\right)^{\alpha-1}\operatorname{d}x}=\frac{\exp_{\kappa}\left[-\left(\frac{x}{\beta}\right)^{\alpha}\right]}{\sqrt{1+\kappa^{2}\left(\frac{x}{\beta}\right)^{2\alpha}}},
\end{equation}
in view of which it is possible to derive the $\kappa$-generalized PDF as the regular derivative of the $\kappa$-deformed exponential distribution, i.e.
\begin{equation}
f\left(x;\alpha,\beta,\kappa\right)=\frac{\operatorname{d}U_{\kappa}\left[\left(\frac{x}{\beta}\right)^\alpha\right]}{\operatorname{d}x}=\frac{\alpha}{\beta}\left(\frac{x}{\beta}\right)^{\alpha-1}
\frac{\exp_{\kappa}\left[-\left(\frac{x}{\beta}\right)^\alpha\right]}{\sqrt{1+\kappa^{2}\left(\frac{x}{\beta}\right)^{2\alpha}}}.
\end{equation}

% Basic properties

\subsubsection{Basic properties}
\label{sec:BasicProperties}

As $\kappa\rightarrow0$, the $\kappa$-generalized distribution tends to the Weibull model with the CDF
\begin{subequations}
\begin{equation}
\lim_{\kappa\to0}F\left(x;\alpha,\beta,\kappa\right)=1-\exp\left[-\left(\frac{x}{\beta}\right)^{\alpha}\right]
\end{equation}
and the PDF
\begin{equation}
\lim_{\kappa\to0}f\left(x;\alpha,\beta,\kappa\right)=\frac{\alpha}{\beta}\left(\frac{x}{\beta}\right)^{\alpha-1}\exp\left[-\left(\frac{x}{\beta}\right)^{\alpha}\right].
\end{equation}
\label{eq:Equation_11}%
\end{subequations}
Consequently, the exponential law is also a special limiting case of the $\kappa$-generalized distribution, since it is a special case of the Weibull with $\alpha=1$.\footnote{The Weibull distribution has received maximum attention in the engineering literature. In physics, it is known as the stretched exponential distribution if $\alpha<1$. In the economic literature the Weibull is probably less prominent, but reference \cite{DAddario1974} noticed its potential for income data\textemdash although it has been used only sporadically as an income distribution (some applications can be found in references \cite{BartelsVanMetelen1975,Bartels1977,EspinguetTerraza1983,McDonald1984,AtodaSurugaTachibanaki1988,BordleyMcDonaldMantrala1996,BrachmannStichTrede1996,TachibanakiSurugaAtoda1997}).}

For $x\rightarrow0^{+}$ the $\kappa$-generalized behaves similarly to the Weibull model \eqref{eq:Equation_11}, whereas for large $x$ it approaches a Pareto distribution with scale $\beta\left(2\kappa\right)^{-\frac{1}{\alpha}}$ and shape $\frac{\alpha}{\kappa}$, i.e.
\begin{subequations}
\begin{equation}
F\left(x;\alpha,\beta,\kappa\right)\underset{x\rightarrow+\infty}{\sim}1-\left[\frac{\beta\left(2\kappa\right)^{-\frac{1}{\alpha}}}{x}\right]^{\frac{\alpha}{\kappa}}
\end{equation}
and
\begin{equation}
f\left(x;\alpha,\beta,\kappa\right)\underset{x\rightarrow+\infty}{\sim}\frac{\frac{\alpha}{\kappa}\left[\beta\left(2\kappa\right)^{-\frac{1}{\alpha}}\right]^{\frac{\alpha}{\kappa}}}{x^{\frac{\alpha}{\kappa}+1}},
\end{equation}
\label{eq:Equation_12}%
\end{subequations}
thus satisfying the weak Pareto law \cite{Mandelbrot1960}.

\begin{figure}[!t]
\centering
\subfigure[]{\includegraphics[width=0.48\textwidth]{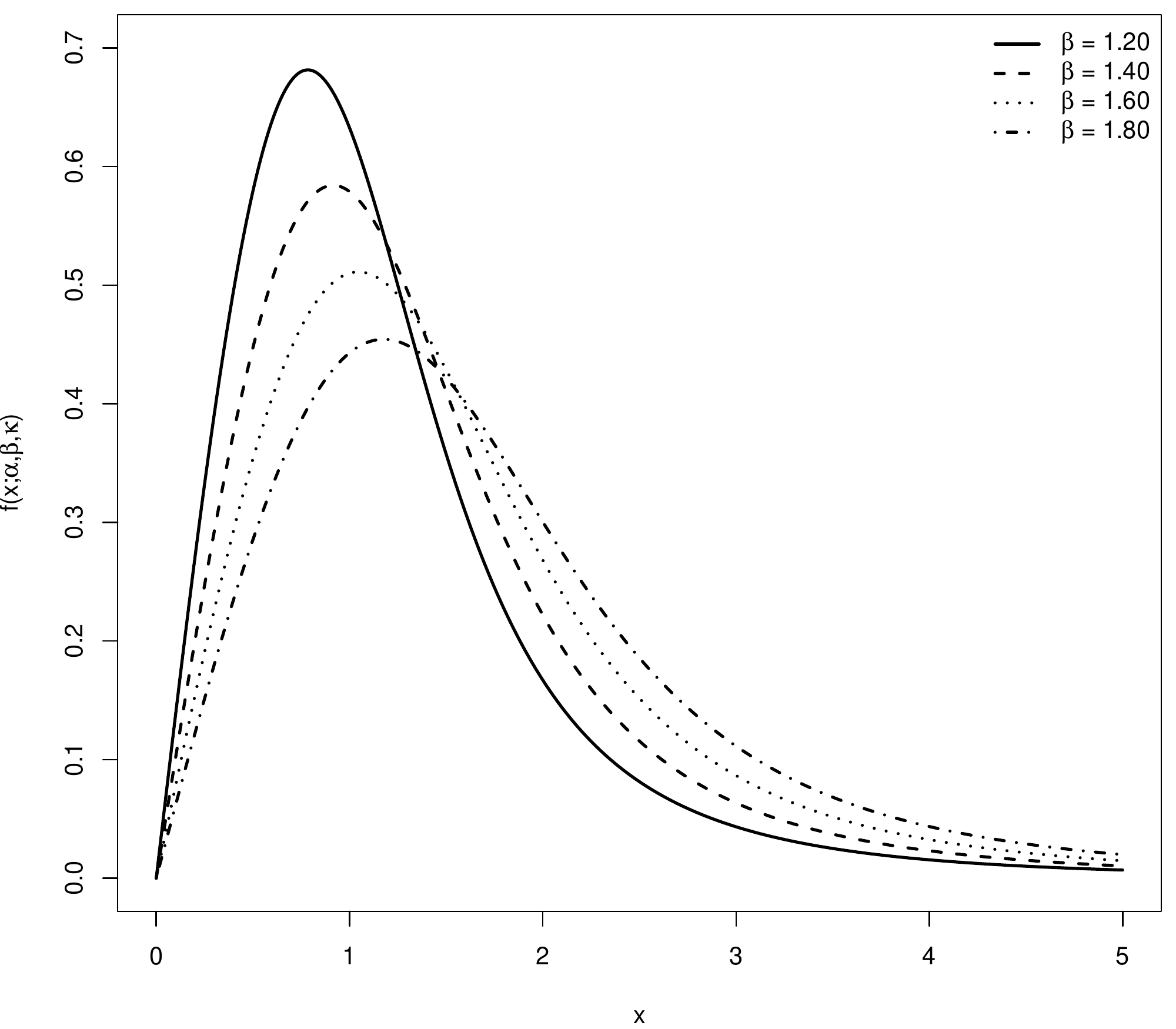}\label{fig:Figure_1_a}}\quad
\subfigure[]{\includegraphics[width=0.48\textwidth]{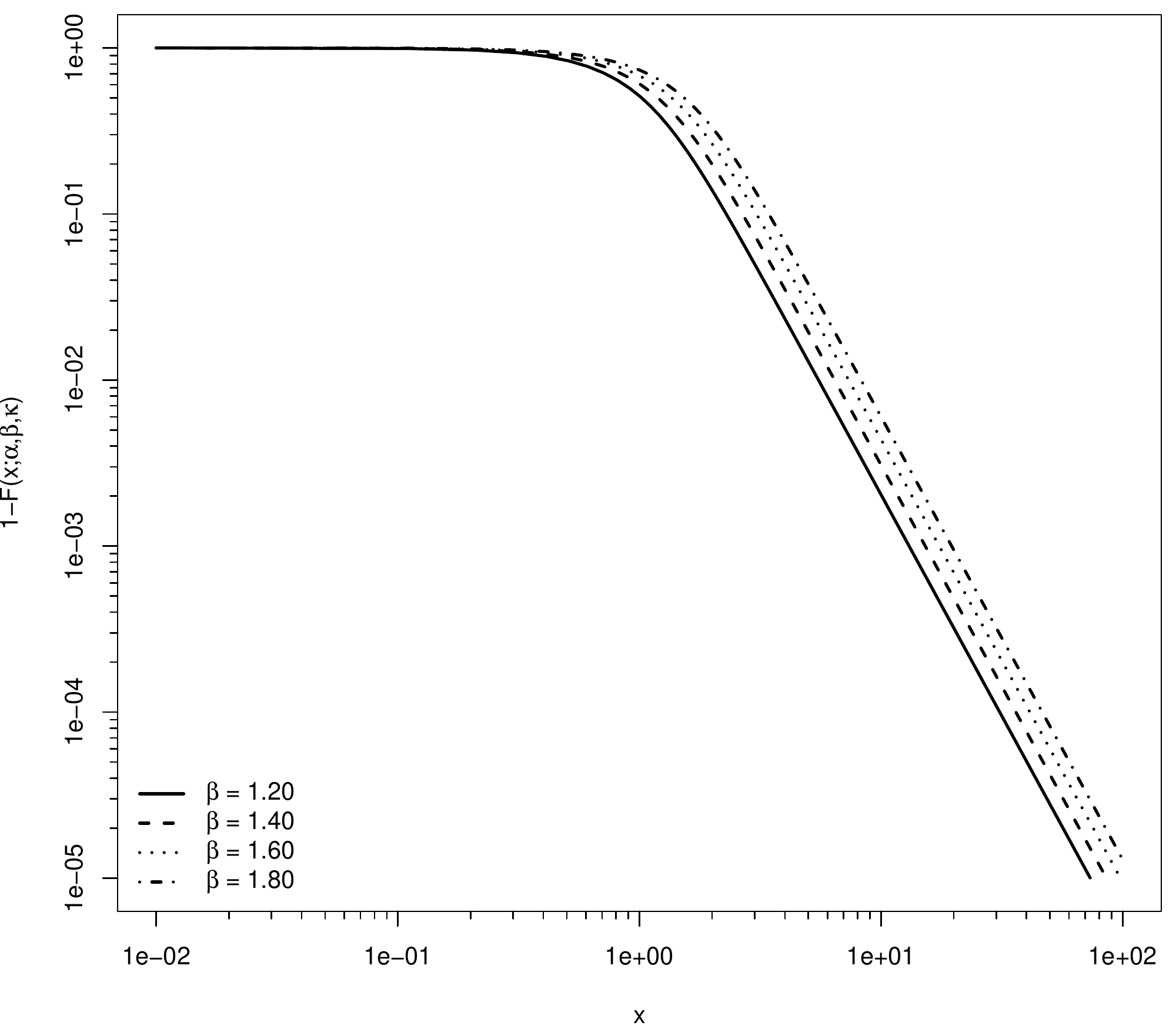}\label{fig:Figure_1_b}}
\caption{Plot of the $\kappa$-generalized PDF \subref{fig:Figure_1_a} and CCDF \subref{fig:Figure_1_b} for some different values of $\beta$ ($=1.20,1.40,1.60,1.80$) and fixed $\alpha$ ($=2.00$) and $\kappa$ ($=0.75$). The CCDF is plotted on doubly-logarithmic axes, which is the standard way of emphasizing the right-tail behavior of a distribution. Notice that the distribution spreads out (concentrates) as the value of $\beta$ increases (decreases).}
\label{fig:Figure_1}
\end{figure}

\begin{figure}[!t]
\centering
\subfigure[]{\includegraphics[width=0.48\textwidth]{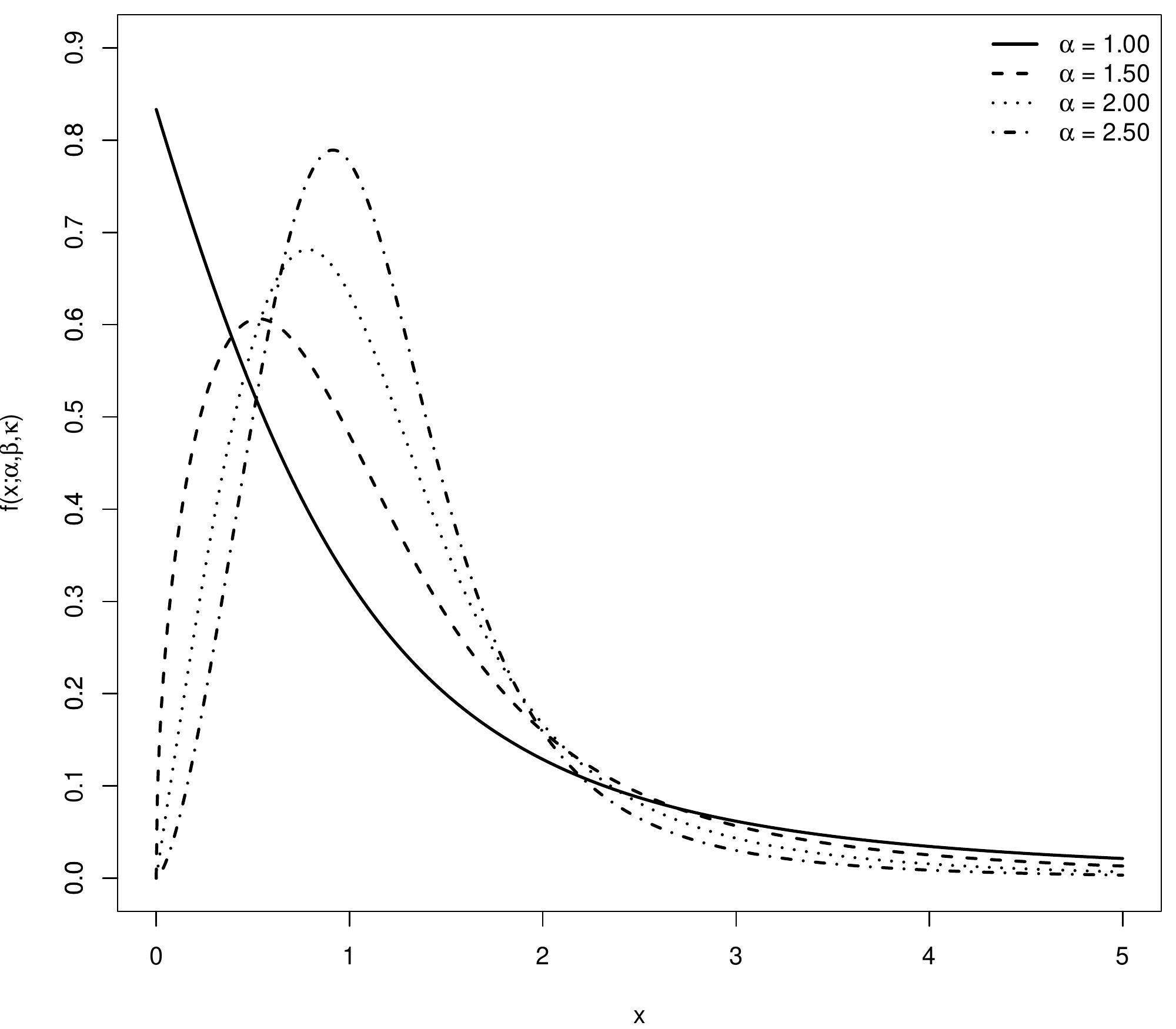}\label{fig:Figure_2_a}}\quad
\subfigure[]{\includegraphics[width=0.48\textwidth]{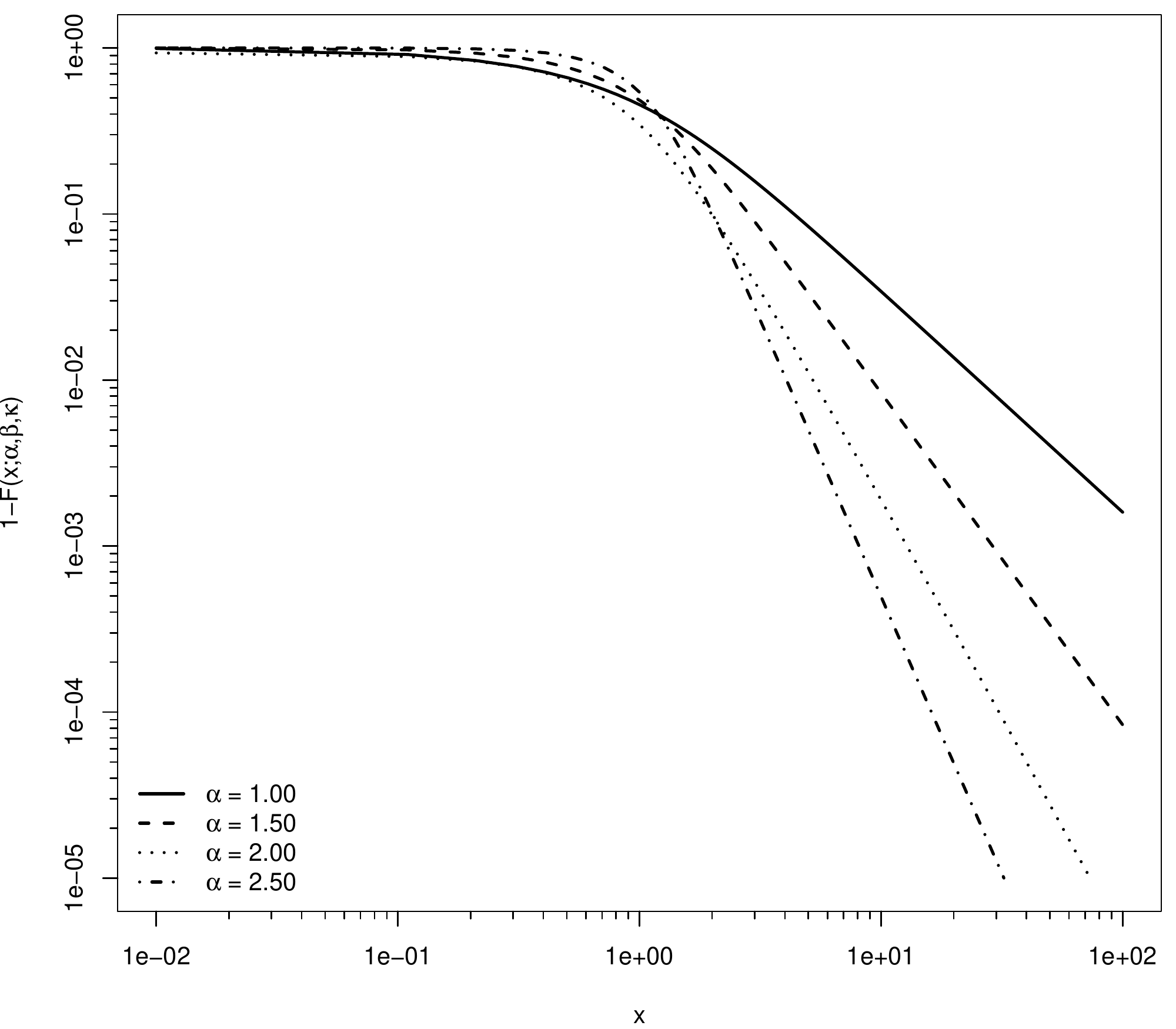}\label{fig:Figure_2_b}}
\caption{Plot of the $\kappa$-generalized PDF \subref{fig:Figure_2_a} and CCDF \subref{fig:Figure_2_b} for some different values of $\alpha$ ($=1.00,1.50,2.00,2.50$) and fixed $\beta$ ($=1.20$) and $\kappa$ ($=0.75$). The CCDF is plotted on doubly-logarithmic axes, which is the standard way of emphasizing the right-tail behavior of a distribution. Notice that the curvature (shape) of the distribution becomes less (more) pronounced when the value of $\alpha$ decreases (increases). The case $\alpha=1.00$ corresponds to the standard exponential distribution.}
\label{fig:Figure_2}
\end{figure}

\begin{figure}[!t]
\centering
\subfigure[]{\includegraphics[width=0.48\textwidth]{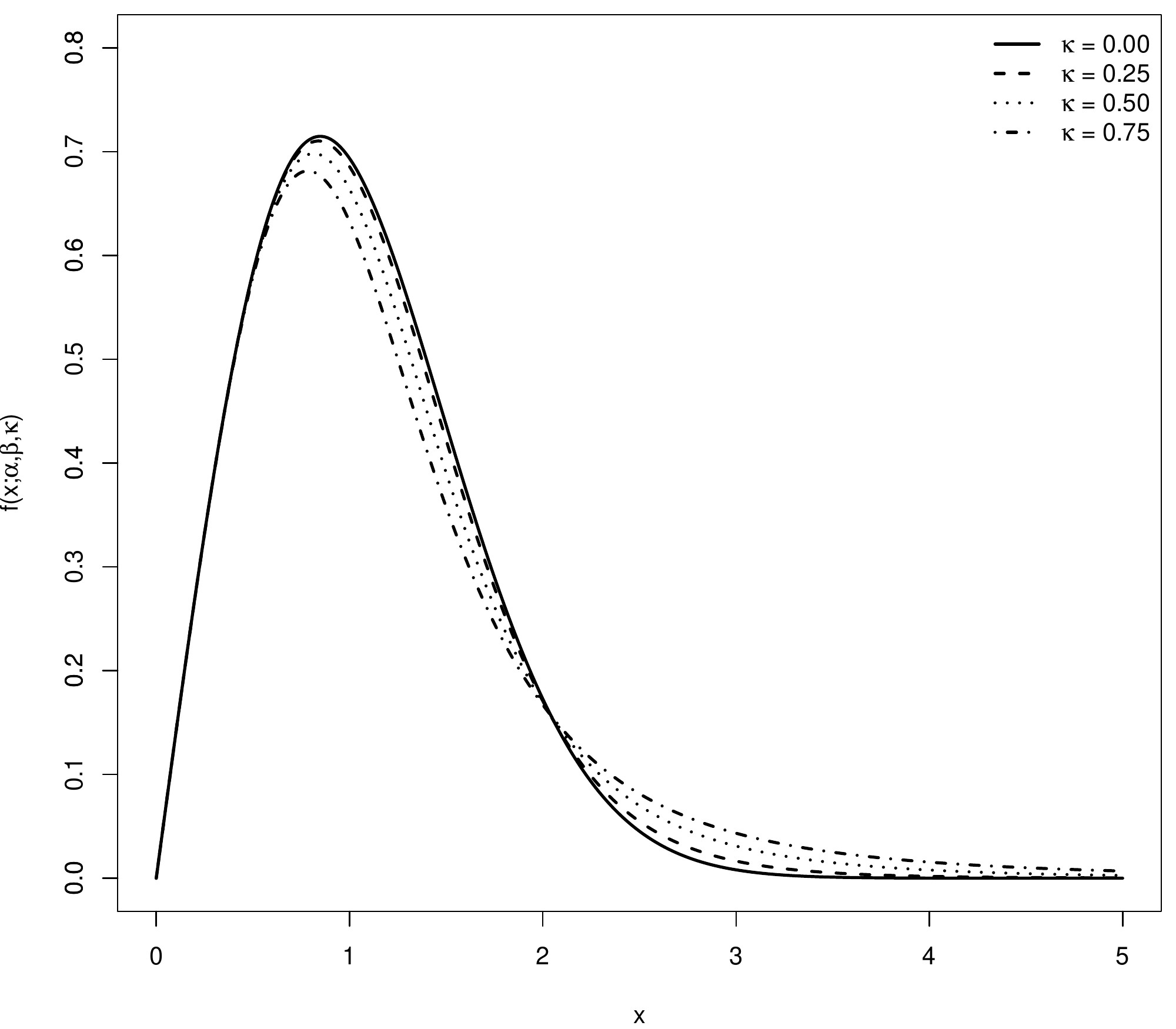}\label{fig:Figure_3_a}}\quad
\subfigure[]{\includegraphics[width=0.48\textwidth]{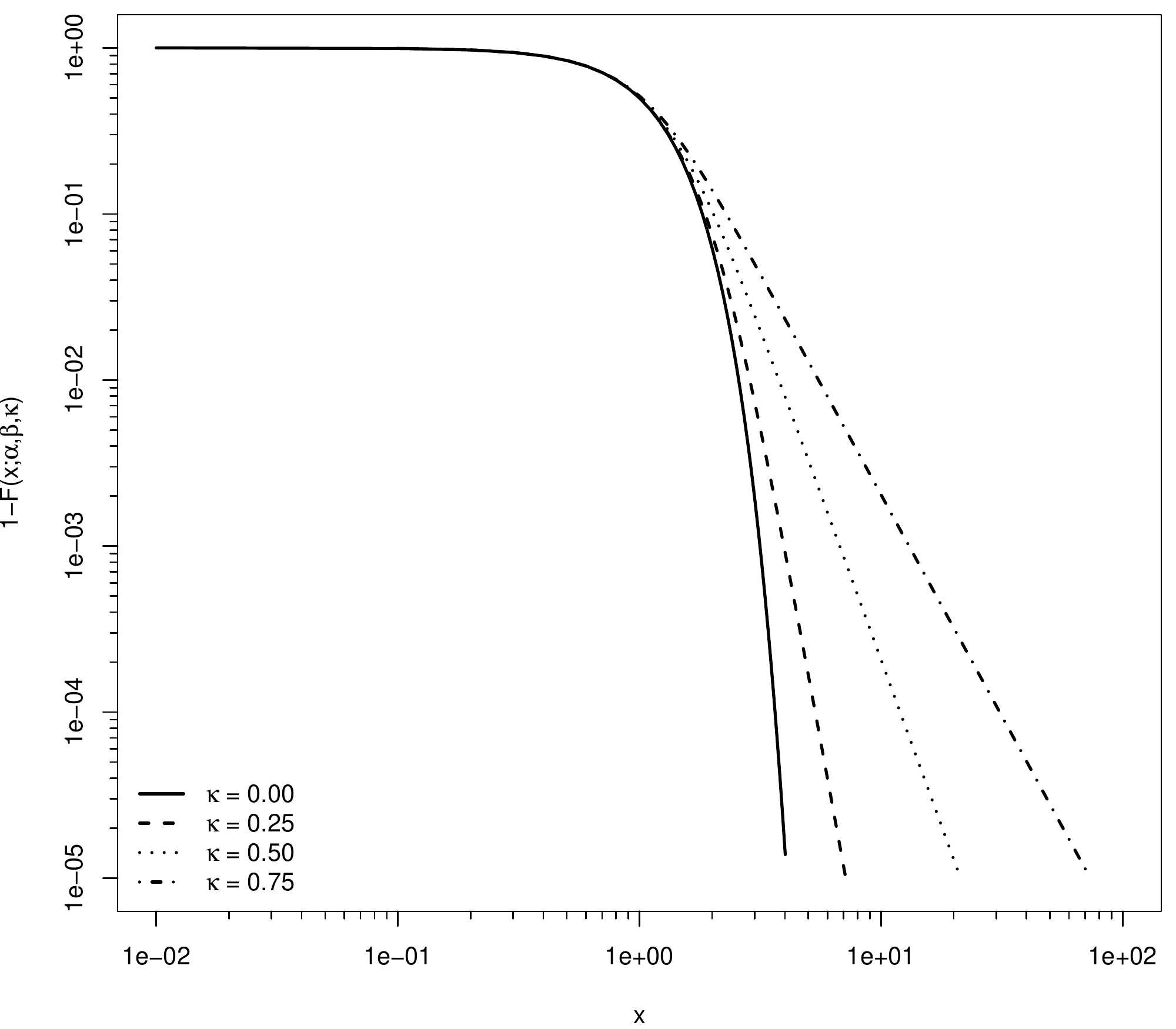}\label{fig:Figure_3_b}}
\caption{Plot of the $\kappa$-generalized PDF \subref{fig:Figure_3_a} and CCDF \subref{fig:Figure_3_b} for some different values of $\kappa$ ($=0.00,0.25,0.50,0.75$) and fixed $\alpha$ ($=2.00$) and $\beta$ ($=1.20$). The CCDF is plotted on doubly-logarithmic axes, which is the standard way of emphasizing the right-tail behavior of a distribution. Notice that the upper tail of the distribution fattens (thins) as the value of $\kappa$ increases (decreases). The case $\kappa=0.00$ corresponds to the Weibull (stretched exponential) distribution.}
\label{fig:Figure_3}
\end{figure}

Figures \ref{fig:Figure_1} to \ref{fig:Figure_3} illustrate the behavior of the $\kappa$-generalized PDF and complementary CDF, $1-F\left(x;\alpha,\beta,\kappa\right)$, for various parameter values. Each graph keeps two parameters constant and varies the remaining one.

The constant $\beta$ is a characteristic scale having the same dimension of income: if $\beta$ is small, then the distribution will be more concentrated around the mode; if $\beta$ is large, then it will be more spread out (Figure \ref{fig:Figure_1}).
%
% Figure 1: Plot of the κ-generalized PDF and CCDF for some different values of β ------------------------
% --------------------------------------------------------------------------------------------------------
%
Since it is related to the income measurement unit, increments (reductions) in the monetary unit lead to a corresponding augmentation (diminishment) in the value of $\beta$ and generate a global increase (decrease) of each and every one of the incomes and, therefore, of the average income.

The exponent $\alpha$, instead, quantifies the curvature (shape) of the distribution, which is less (more) pronounced for lower (higher) values of the parameter, as seen in Figure \ref{fig:Figure_2}.
%
% Figure 2: Plot of the κ-generalized PDF and CCDF for some different values of α ------------------------
% --------------------------------------------------------------------------------------------------------
%
It should be noted that for $\alpha\leq1$ the density exhibits a pole at the origin, whereas for $\alpha>1$ there exists an interior mode. In the latter case, this mode is at
\begin{equation}
x_{\mathrm{mode}}=\beta\left[\frac{\alpha^{2}+2\kappa^{2}\left(\alpha-1\right)}{2\kappa^{2}\left(\alpha^{2}-\kappa^{2}\right)}\right]^{\frac{1}{2\alpha}}\left\{\sqrt{1+\frac{4\kappa^{2}\left(\alpha^{2}-\kappa^{2}\right)\left(\alpha-1\right)^{2}}{\left[\alpha^{2}+2\kappa^{2}\left(\alpha-1\right)\right]^{2}}}-1\right\}^{\frac{1}{2\alpha}}.
\end{equation}
This built-in flexibility is an attractive feature in that the model can approximate income distributions, which are usually uni-modal, and wealth distributions, which are zero-modal (see Section \ref{sec:TheKappaGeneralizedMixtureModelForTheSizeDistributionOfWealth}).

Finally, as Figure \ref{fig:Figure_3} shows, the deformation parameter $\kappa$ measures the fatness of the upper tail: the larger (smaller) its magnitude, the fatter (thinner) the tail.\footnote{Differently than $\beta$, $\alpha$ and $\kappa$ are scale-free (dimensionless) parameters: changes in the monetary unit yield $\beta$ modifications but leave $\alpha$ and $\kappa$ invariant.}
%
% Figure 3: Plot of the κ-generalized PDF and CCDF for some different values of κ ------------------------
% --------------------------------------------------------------------------------------------------------

The $\kappa$-generalized admits a closed-form expression for the quantile function,
\begin{equation}
F^{-1}\left(u;\alpha,\beta,\kappa\right)=\beta\left[\ln_{\kappa}\left(\frac{1}{1-u}\right)\right]^{\frac{1}{\alpha}},\quad 0<u<1,
\label{eq:Equation_14}
\end{equation}
where $\ln_{\kappa}\left(\cdot\right)$ denotes the deformed logarithmic function defined as the inverse function of \eqref{eq:Equation_A4}, namely $\ln_{\kappa}\left[\exp_{\kappa}\left(x\right)\right]=\exp_{\kappa}\left[\ln_{\kappa}\left(x\right)\right]=x$, and reads as
\begin{equation}
\ln_{\kappa}\left(x\right)=\frac{x^{\kappa}-x^{-\kappa}}{2\kappa},\;\;\,x\in\mathbb{R}_{+}.
\end{equation}
Hence random numbers from a $\kappa$-generalized distribution can be easily generated via the inversion method.

The $r$\textsuperscript{th} moment exists for $-\alpha<r<\frac{\alpha}{\kappa}$ and equals
\begin{equation}
E\left(X^{r}\right)=\beta^{r}\left(2\kappa\right)^{-\frac{r}{\alpha}}\frac{\Gamma\left(1+\frac{r}{\alpha}\right)}{1+\frac{r}{\alpha}\kappa}\frac{\Gamma\left(\frac{1}{2\kappa}-\frac{r}{2\alpha}\right)}{\Gamma\left(\frac{1}{2\kappa}+\frac{r}{2\alpha}\right)},
\label{eq:Equation_15}
\end{equation}
where $\Gamma\left(\cdot\right)$ denotes the gamma function. Specifically,
\begin{equation}
E\left(X\right)=m=\beta\left(2\kappa\right)^{-\frac{1}{\alpha}}\frac{\Gamma\left(1+\frac{1}{\alpha}\right)}{1+\frac{\kappa}{\alpha}}\frac{\Gamma\left(\frac{1}{2\kappa}-\frac{1}{2\alpha}\right)}{\Gamma\left(\frac{1}{2\kappa}+\frac{1}{2\alpha}\right)}
\end{equation}
is the mean of the distribution and
\begin{equation}
\mathrm{Var}\left(X\right)=\beta^{2}\left(2\kappa\right)^{-\frac{2}{\alpha}}\left\{\frac{\Gamma\left(1+\frac{2}{\alpha}\right)}{1+2\frac{\kappa}{\alpha}}\frac{\Gamma\left(\frac{1}{2\kappa}-\frac{1}{\alpha}\right)}{\Gamma\left(\frac{1}{2\kappa}+\frac{1}{\alpha}\right)}-\left[\frac{\Gamma\left(1+\frac{1}{\alpha}\right)}{1+\frac{\kappa}{\alpha}}\frac{\Gamma\left(\frac{1}{2\kappa}-\frac{1}{2\alpha}\right)}{\Gamma\left(\frac{1}{2\kappa}+\frac{1}{2\alpha}\right)}\right]^{2}\right\}
\end{equation}
is the variance.

% Measuring income inequality using the κ-generalized distribution

\subsubsection{Measuring income inequality using the $\boldsymbol{\kappa}$-generalized distribution}
\label{sec:MeasuringIncomeInequalityUsingTheKappaGeneralizedDistribution}

The most widely used tool for analyzing and visualizing income inequality is the Lorenz curve \cite{Lorenz1905}, and several indices of income inequality are directly related to this curve, most notably
the Gini index \cite{Gini1914}.

Since the quantile function of the $\kappa$-generalized distribution is available in closed form, its normalized integral, the Lorenz curve
\begin{equation}
L\left(u\right)=\frac{1}{m}\int\limits^{u}_{0}F^{-1}\left(t\right)\operatorname{d}t,\quad u\in\left[0,1\right],
\end{equation}
is also of a comparatively simple form, namely \cite{Okamoto2013}
\begin{equation}
L\left(u\right)=I_{X}\left(1+\frac{1}{\alpha},\frac{1}{2\kappa}-\frac{1}{2\alpha}\right),\quad X=1-\left(1-u\right)^{2\kappa},
\end{equation}
where $I_{X}\left(\cdot,\cdot\right)$ is the regularized incomplete beta function defined in terms of the incomplete beta function and the complete beta function\textemdash that is, $I_{X}\left(\cdot,\cdot\right)=\frac{B_{X}\left(\cdot,\cdot\right)}{B\left(\cdot,\cdot\right)}$. The curve exists if and only if $\frac{\alpha}{\kappa}>1$.

\begin{figure}[!t]
\centering
\subfigure[]{\includegraphics[width=0.48\textwidth]{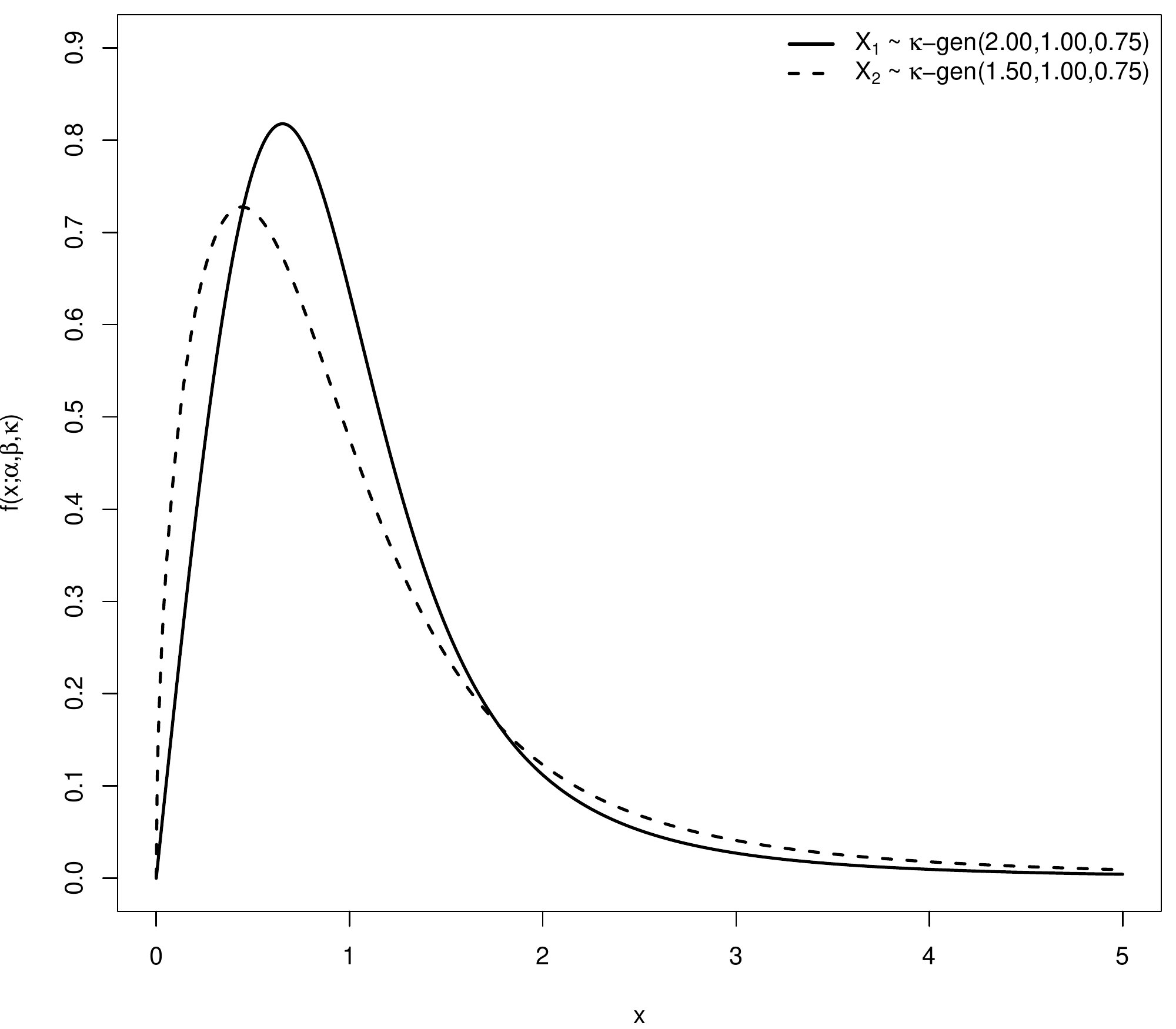}\label{fig:Figure_4_a}}\quad
\subfigure[]{\includegraphics[width=0.48\textwidth]{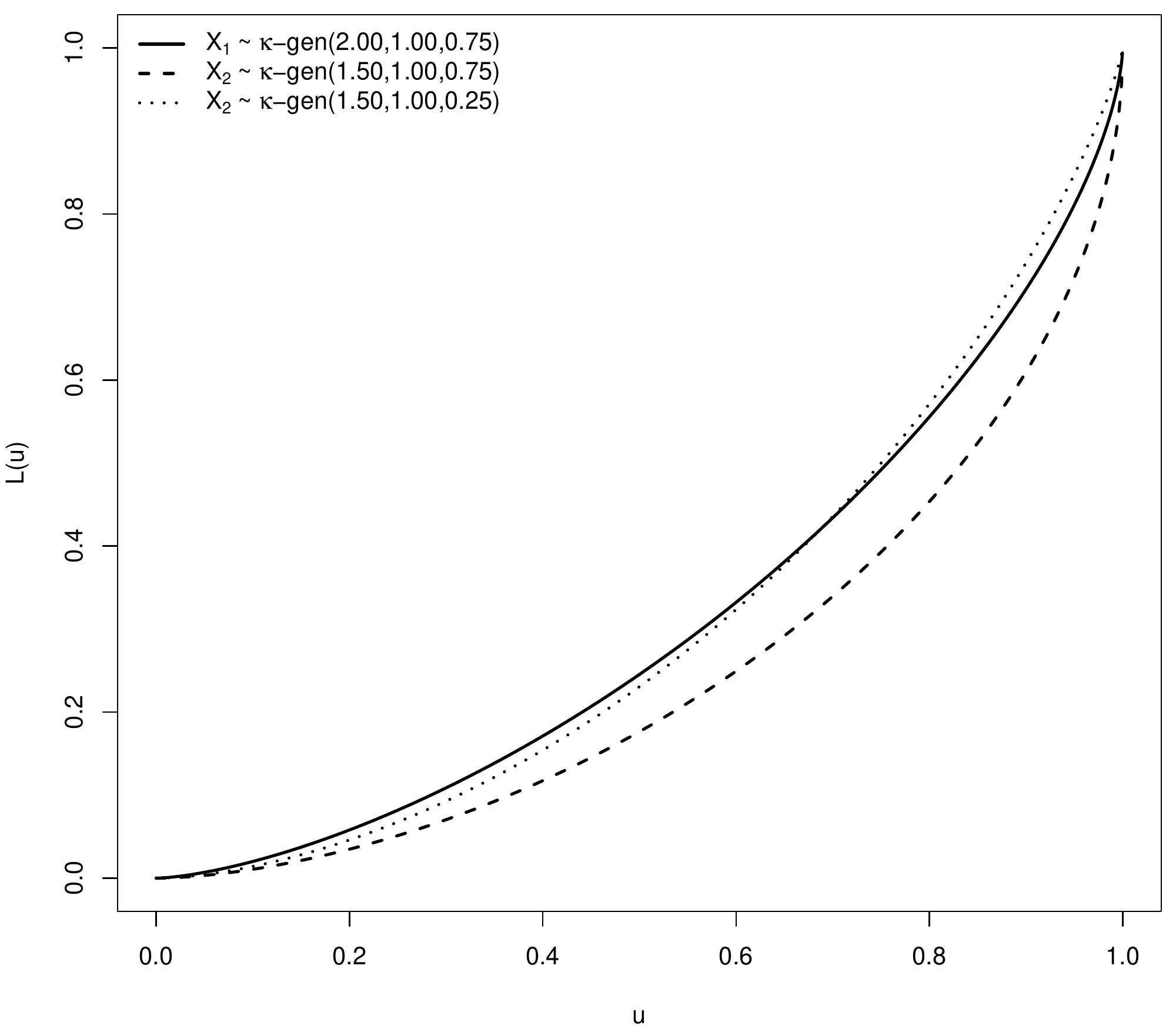}\label{fig:Figure_4_b}}
\caption{Tails \subref{fig:Figure_4_a} and Lorenz curves \subref{fig:Figure_4_b} for two $\kappa$-generalized distributions. As shown by the dotted line, the Lorenz curves intersect if the parameters are not chosen according to Equation \eqref{eq:Equation_20}.}
\label{fig:Figure_4}
\end{figure}

For the comparison of estimated income distributions it is often of interest to know the parameter constellations for which Lorenz curves do or do not intersect. A complete analytical characterization for the $\kappa$-generalized distribution was obtained by \cite{ClementiGallegatiKaniadakis2010}. Suppose $X_{i}\sim\kappa\text{-gen}\left(\alpha_{i},\beta_{i},\kappa_{i}\right)$, $i=1,2$. The necessary and sufficient conditions for which the Lorenz curves of $X_{1}$ and $X_{2}$ do not intersect and we have $X_{1}\leq_{L}X_{2}$\textemdash i.e. the Lorenz curve of $X_{1}$ lies nowhere below that of $X_{2}$ for all $u\in\left[0,1\right]$ and consequently $X_{1}$ exhibits less inequality than $X_{2}$ in the Lorenz sense\textemdash are
\begin{equation}
X_{1}\leq_{L}X_{2}\Longleftrightarrow\alpha_{1}\geq\alpha_{2}\quad\text{and}\quad\frac{\alpha_{1}}{\kappa_{1}}\geq\frac{\alpha_{2}}{\kappa_{2}}.
\label{eq:Equation_20}
\end{equation}
Figure \ref{fig:Figure_4} provides an illustration of \eqref{eq:Equation_20}, showing that the less unequal distribution (in the Lorenz sense) always exhibits lighter tails.
%
% Figure 4: Tails and Lorenz curves for two κ-generalized distributions ----------------------------------
% --------------------------------------------------------------------------------------------------------

As regards scalar measures of inequality, the most widely used of all such indices, the Gini coefficient, takes the form
\begin{equation}
G=1-\frac{2\alpha+2\kappa}{2\alpha+\kappa}\frac{\Gamma\left(\frac{1}{\kappa}-\frac{1}{2\alpha}\right)}{\Gamma\left(\frac{1}{\kappa}+\frac{1}{2\alpha}\right)}\frac{\Gamma\left(\frac{1}{2\kappa}+\frac{1}{2\alpha}\right)}{\Gamma\left(\frac{1}{2\kappa}-\frac{1}{2\alpha}\right)}.
\label{eq:Equation_21}
\end{equation}
Using Stirling approximation for the gamma function, $\Gamma\left(z\right)\approx\sqrt{2\pi}z^{z-\frac{1}{2}}\exp\left(-z\right)$, and taking the limit as $\kappa\rightarrow0$ in Equation \eqref{eq:Equation_21}, one arrives after some simplification at $G=1-2^{-\frac{1}{\alpha}}$, which is the explicit form of the Gini coefficient for the Weibull distribution \cite{KleiberKotz2003}. Since the exponential distribution is a special case of the Weibull with shape parameter equal to 1, one directly determines that for $\kappa\to0$ and $\alpha=1$ the Gini coefficient of exponentially distributed incomes is equal to one half \cite{DragulescuYakovenko2001}.\footnote{As it is clear from Figure \ref{fig:Figure_4_a}, the region around the origin of the $\kappa$-generalized distribution is governed by $\alpha$, the upper tail by both $\alpha$ and $\kappa$. More precisely, increasing $\kappa$ leads to a thicker upper tail, whereas increasing $\alpha$ lowers both the tails and yields a greater concentration of probability mass around the peak of the distribution. Hence, the $\alpha$ and $\kappa$ modifications affect the shape of the distribution and are related to the inequality of the income distribution\textemdash as they enter into several inequality measures that can be derived from the $\kappa$-generalized model of income distribution. However, the influence of each parameter on these measures is different. For instance, reference \cite{ClementiGallegatiKaniadakis2010} obtained a first interpretation of $\kappa$ as an ``inequality'' parameter and $\alpha$ as an ``equality'' parameter after verifying that the Gini coefficient \eqref{eq:Equation_21} increases with the former and decreases with the latter. This interpretation has been further corroborated by \cite{ClementiGallegati2016}.}

Although commonly used, the Gini is but one of many measures of inequality that are available, and it incorporates particular assumptions about the way in which income differences in different parts of the distribution are summarized\textemdash it tends to be most sensitive to transfers around the middle
of the income distribution and least sensitive to transfers among the very rich or the very poor \cite{Allison1978}. In other forms of income distribution research, the generalized entropy class of inequality measures \cite{Cowell1980a,Cowell1980b,Shorrocks1980,CowellKuga1981a,CowellKuga1981b} provides a full range of bottom- to top-sensitive indices that are widely used by analysts to assess inequality in different parts of the income distribution. In terms of the $\kappa$-generalized parameters, this class of inequality indices is expressed as
\begin{equation}
GE\left(\theta\right)=\frac{1}{\theta^{2}-\theta}\left\{\left(\frac{\beta}{m}\right)^{\theta}\left[\frac{\left(2\kappa\right)^{-\frac{\theta}{\alpha}}}{1+\frac{\theta}{\alpha}\kappa}\frac{\Gamma\left(\frac{1}{2\kappa}-\frac{\theta}{2\alpha}\right)}{\Gamma\left(\frac{1}{2\kappa}+\frac{\theta}{2\alpha}\right)}\Gamma\left(1+\frac{\theta}{\alpha}\right)\right]-1\right\},\;\;\,\theta\neq0,1.
\label{eq:Equation_22}
\end{equation}
Equation \eqref{eq:Equation_22} defines a class because the index $GE\left(\theta\right)$ assumes different forms depending on the value assigned to $\theta$, the parameter that characterizes the sensitivity of $GE\left(\theta\right)$ to income differences in different parts of the distribution\textemdash the more positive that $\theta$ is, the more sensitive is $GE\left(\theta\right)$ to income differences at the top of the distribution; the more negative that $\theta$ is, the more sensitive is $GE\left(\theta\right)$ to income differences at the bottom of the distribution. In applied work, two limiting cases of \eqref{eq:Equation_22} are of particular interest for inequality measurement: the mean logarithmic deviation index
\begin{equation}
MLD=\lim_{\theta\rightarrow0}GE\left(\theta\right)=\frac{1}{\alpha}\left[\gamma+\psi\left(\frac{1}{2\kappa}\right)+\ln\left(2\kappa\right)-\alpha\ln\left(\frac{\beta}{m}\right)+\kappa\right],
\label{eq:Equation_23}
\end{equation}
where $\gamma=-\psi\left(1\right)$ is the Euler-Mascheroni constant and $\psi\left(z\right)=\Gamma^{'}\left(z\right)/\Gamma\left(z\right)$ is the digamma function, and the Theil index \cite{Theil1967}
\begin{equation}
\begin{split}
T=\lim_{\theta\rightarrow1}GE\left(\theta\right)=&\frac{1}{\alpha}\left[\psi\left(1+\frac{1}{\alpha}\right)-\frac{1}{2}\psi\left(\frac{1}{2\kappa}-\frac{1}{2\alpha}\right)-\frac{1}{2}\psi\left(\frac{1}{2\kappa}+\frac{1}{2\alpha}\right)\right.\\
&\left.-\ln\left(2\kappa\right)+\alpha\ln\left(\frac{\beta}{m}\right)-\frac{\alpha\kappa}{\alpha+\kappa}\right].
\end{split}
\label{eq:Equation_24}
\end{equation}
Expression for each index other than for the cases \eqref{eq:Equation_23} and \eqref{eq:Equation_24} can be derived by straightforward substitution.\footnote{Further income inequality measures are presented in \cite{ClementiGallegati2016}. Notice that all the measures considered here are functions of the distribution moments, whose existence depends on some conditions guaranteeing the convergence of relevant integrals. As a matter of example, the Gini coefficient \eqref{eq:Equation_21} exists provided the mean of the distribution $m=\int_{0}^{\infty}xf\left(x;\alpha,\beta,\kappa\right)\operatorname{d}x$ converges, which is true when $\frac{\alpha}{\kappa}>1$. As shown by \cite{Kleiber1997}, the problem of existence of popular inequality measures is common to various parametric models of income distribution.}

% Estimation and inference

\subsubsection{Estimation and inference}
\label{sec:EstimationAndInference}

Parameter estimation for the $\kappa$-generalized distribution can be performed using the maximum likelihood method, which yields estimators with good statistical properties \cite{Rao1973,Ghosh1994}. Assuming that all sample observations $\mathbf{x}=\left\{x_{1},\ldots,x_{n}\right\}$ are independent, the likelihood function is
\begin{equation}
L\left(\mathbf{x};\boldsymbol{\theta}\right)=\prod\limits^{n}_{i=1}f\left(x_{i};\boldsymbol{\theta}\right)^{w_{i}}=\prod\limits^{n}_{i=1}\left\{\frac{\alpha}{\beta}\left(\frac{x_{i}}{\beta}\right)^{\alpha-1}\frac{\exp_{\kappa}\left[-\left(x_{i}/\beta\right)^{\alpha}\right]}{\sqrt{1+\kappa^{2}\left(x_{i}/\beta\right)^{2\alpha}}}\right\}^{w_{i}},
\end{equation}
where $f\left(x_{i};\boldsymbol{\theta}\right)$ is the probability distribution function, $\boldsymbol{\theta}=\left\{\alpha,\beta,\kappa\right\}$ is the unknown parameters vector, $w_{i}$ is the weight of $i$\textsuperscript{th} observation and $n$ is the sample size. This leads to the problem of solving the partial derivatives of the log-likelihood function
\begin{equation}
l\left(\mathbf{x};\boldsymbol{\theta}\right)=\ln\left[L\left(\mathbf{x};\boldsymbol{\theta}\right)\right]=\sum_{i=1}^{n}w_{i}\ln\left[f\left(x_{i};\boldsymbol{\theta}\right)\right]
\label{eq:Equation_26}
\end{equation}
with respect to $\alpha$, $\beta$ and $\kappa$, which corresponds to finding the solution of the following non-linear system of equations
\begin{equation}
\sum_{i=1}^{n}w_{i}\frac{\partial}{\partial\alpha}\ln\left[f\left(x_{i};\boldsymbol{\theta}\right)\right]=0,
\end{equation}
\begin{equation}
\sum_{i=1}^{n}w_{i}\frac{\partial}{\partial\beta}\ln\left[f\left(x_{i};\boldsymbol{\theta}\right)\right]=0,
\end{equation}
\begin{equation}
\sum_{i=1}^{n}w_{i}\frac{\partial}{\partial\kappa}\ln\left[f\left(x_{i};\boldsymbol{\theta}\right)\right]=0.
\end{equation}
However, obtaining explicit expressions for the maximum likelihood estimators of the three parameters by resolution of the above equations is difficult, making direct analytical solutions intractable. The derivation of any of these estimators, therefore, generally involves the use of numerical optimization algorithms.\footnote{As already noted in Section \ref{sec:Definitions}, references \cite{ClementiGallegatiKaniadakis2007,ClementiDiMatteoGallegatiKaniadakis2008,ClementiGallegatiKaniadakis2009} use a slightly different parametrization of the $\kappa$-generalized PDF, namely $f\left(z;\alpha,\lambda,\kappa\right)=\alpha\lambda z^{\alpha-1}\frac{\exp_{\kappa}\left(-\lambda z^{\alpha}\right)}{\sqrt{1+\kappa^{2}\lambda^{2}z^{2\alpha}}}$, where $\lambda=\beta^{-\alpha}$ and $z$ is defined as $z=\frac{x}{m}$, being $x$ the absolute income and $m$ its mean value. Taking into account the meaning of the variable $z$, the mean value results to be equal to unity, i.e. $z=\int_{0}^{\infty}zf\left(z;\alpha,\lambda,\kappa\right)\operatorname{d}z=1$. The latter relationship permits to express the scale parameter $\lambda$ as a function of the shape parameters $\alpha$ and $\kappa$, obtaining $\lambda=\frac{1}{2\kappa}\left[\frac{\Gamma\left(\frac{1}{\alpha}\right)}{\kappa+\alpha}\frac{\Gamma\left(\frac{1}{2\kappa}-\frac{1}{2\alpha}\right)}{\Gamma\left(\frac{1}{2\kappa}+\frac{1}{2\alpha}\right)}\right]^{\alpha}$. In this way, the problem to determine the values of the free parameters $\left\{\alpha,\lambda,\kappa\right\}$ from the empirical data reduces to a two-parameter $\left\{\alpha,\kappa\right\}$ fitting problem. To find the parameter values such that the negative of $l\left(\mathbf{z};\boldsymbol{\theta}\right)$ is minimized, one can use the constrained maximum likelihood estimation method \cite{Schoenberg1997}, which solves the general maximum log-likelihood problem \eqref{eq:Equation_26} subject to the non-linear equality constraint given by $\lambda$ and bounds $\alpha,\lambda>0$ and $\kappa\in\left[0,1\right)$.}

% The κ-generalized mixture model for the size distribution of wealth ------------------------------------

\subsection{The $\boldsymbol{\kappa}$-generalized mixture model for the size distribution of wealth}
\label{sec:TheKappaGeneralizedMixtureModelForTheSizeDistributionOfWealth}

The $\kappa$-generalized distribution was also successfully used in a three-component mixture model for analyzing the singularities of survey data on \textit{net} wealth, i.e. the value of gross wealth minus total debt, which present highly significant frequencies of households or individuals with null and/or negative wealth \cite{ClementiGallegatiKaniadakis2012b,ClementiGallegati2016}. The support of the $\kappa$-generalized mixture model for net wealth distribution is the real line $\mathbb{R}=\left(-\infty,\infty\right)$, thus allowing to fit the subset of economic units with nil and negative net worth. Furthermore, it contains as a particular case the $\kappa$-generalized model for income distributions.

More in detail, the $\kappa$-generalized model of net wealth distribution is a mixture (or a convex combination) of an atomic and two continuous distributions. The atomic distribution concentrates its unit mass of economic agents at zero, and therefore accounts for the economic units with null net wealth. The continuous distribution accounting for the negative net wealth observations is given by a Weibull function. It has a fast left tail convergence to zero, and therefore it has finite moments of all orders. The other continuous distribution, specified as the $\kappa$-generalized model \eqref{eq:Equation_2}, accounts for the positive values of net wealth and presents a heavy right tail, thus having a small number of finite moments of positive order. This different behavior at the two tails of the distribution stems form the fact that, unlike the right tail of income and (gross or net) wealth distributions\textemdash which tend slowly to zero when income and wealth tend to infinity\textemdash the distribution of the negative values (left tail) of net wealth tends very fast to zero when the variable tends to minus infinity, since economic units face a short-term challenge of either moving out of the negative range of net wealth or bankruptcy.

% Model specification

\subsubsection{Model specification}
\label{sec:ModelSpecification}

Formally, the $\kappa$-generalized model of net wealth distribution as a mixture of an atomic and two continuous distributions is specified as
\begin{equation}
f\left(w\right)=\sum^{3}_{i=1}\theta_{i}f_{i}\left(w\right),\;\;\, -\infty<w<\infty,\;\;\, \theta_{i}\geq0,\;\;\, \sum_{i}\theta_{i}=1,
\label{eq:Equation_30}
\end{equation}
where $w$ denotes the wealth variable and $\theta_{i}$, $i=1,\ldots,3$, are the mixture proportions. The two-parameter Weibull density
\begin{equation}
f_{1}\left(w\right)=\frac{s}{\lambda}\left(\frac{\left|w\right|}{\lambda}\right)^{s-1}\exp\left[-\left(\frac{\left|w\right|}{\lambda}\right)^{s}\right],\;\;\, w<0,\;\;\, s,\lambda>0
\end{equation}
describes the distribution of economic units with negative net wealth, while the null net wealth observations are accounted for by a distribution that concentrates its unit mass at $w=0$, i.e.
\begin{equation}
f_{2}\left(0\right)=1.
\end{equation}
The other continuous distribution, $f_{3}\left(w\right)$, accounts for the positive values of net wealth and is specified by the three-parameter $\kappa$-generalized density
\begin{equation}
f_{3}\left(w\right)=\frac{\alpha}{\beta}\left(\frac{w}{\beta}\right)^{\alpha-1}\frac{\exp_{\kappa}\left[-\left(w/\beta\right)^{\alpha}\right]}{\sqrt{1+\kappa^{2}\left(w/\beta\right)^{2\alpha}}},\;\;\, w>0,\;\;\,\alpha,\beta>0,\;\;\,\kappa\in\left[0,1\right).
\end{equation}

The corresponding cumulative distribution function reads
\begin{equation}
F\left(w\right)=\theta_{1}F_{1}\left(w\right)+\theta_{2}F_{2}\left(w\right)+\theta_{3}F_{3}\left(w\right),
\end{equation}
with
\begin{subequations}
\begin{align}
F_{1}\left(w\right)&=
\begin{cases}
\exp\left[-\left(\frac{\left|w\right|}{\lambda}\right)^{s}\right]&\text{if $w<0$},\\
1&\text{if $w\geq0$};
\end{cases}
\\
F_{2}\left(w\right)&=
\begin{cases}
0&\text{if $w<0$},\\
1&\text{if $w\geq0$};
\end{cases}
\\
F_{3}\left(w\right)&=
\begin{cases}
0&\text{if $w\leq0$},\\
1-\exp_{\kappa}\left[-\left(w/\beta\right)^{\alpha}\right]&\text{if $w>0$}.
\end{cases}
\end{align}%
\end{subequations}
It follows easily that
\begin{equation}
F\left(w\right)=
\begin{cases}
\theta_{1}\exp\left[-\left(\frac{\left|w\right|}{\lambda}\right)^{s}\right]&\text{if $w<0$},\\
\rho&\text{if $w=0$},\\
\rho+\left(1-\rho\right)\left\{1-\exp_{\kappa}\left[-\left(w/\beta\right)^{\alpha}\right]\right\}&\text{if $w>0$},
\end{cases}
\label{eq:Equation_36}
\end{equation}
where $\rho=\theta_{1}+\theta_{2}$ and $1-\rho=\theta_{3}$. It can be verified that when $\theta_{1}=\theta_{2}=0$ (hence $\rho=0$) and $\theta_{3}=1-\rho=1$, the $\kappa$-generalized distribution function \eqref{eq:Equation_1} is recovered.

% Moments of the k-generalized mixture model for net wealth distribution

\subsubsection{Moments of the $\kappa$-generalized mixture model for net wealth distribution}
\label{sec:MomentsOfTheKappaGeneralizedMixtureModelForNetWealthDistribution}

From \eqref{eq:Equation_30}, the $r$\textsuperscript{th}-order moment about the origin is
\begin{equation}
E\left(W^{r}\right)=\int\limits^{\infty}_{-\infty}w^{r}f\left(w\right)\operatorname{d}w=\theta_{1}E_{1}\left(W^{r}\right)+\theta_{2}E_{2}\left(W^{r}\right)+\theta_{3}E_{3}\left(W^{r}\right),
\end{equation}
with
\begin{subequations}
\begin{align}
E_{1}\left(W^{r}\right)&=(-1)^{r}\lambda^{r}\Gamma\left(1+\frac{r}{s}\right),
\\
E_{2}\left(W^{r}\right)&=0,
\\
E_{3}\left(W^{r}\right)&=\beta^{r}\left(2\kappa\right)^{-\frac{r}{\alpha}}\frac{\Gamma\left(1+\frac{r}{\alpha}\right)}{1+\frac{r}{\alpha}\kappa}\frac{\Gamma\left(\frac{1}{2\kappa}-\frac{r}{2\alpha}\right)}{\Gamma\left(\frac{1}{2\kappa}+\frac{r}{2\alpha}\right)}.
\label{eq:Equation_38_c}
\end{align}%
\end{subequations}
Specifically, the mean net wealth equals
\begin{equation}
E\left(W\right)=m=-\theta_{1}\lambda\Gamma\left(1+\frac{1}{s}\right)+\theta_{3}E_{3}\left(W\right),
\end{equation}
where $E_{3}\left(W\right)$ is given by Equation \eqref{eq:Equation_38_c} with $r=1$ [compare with \eqref{eq:Equation_15}].

% The Lorenz curve and the Gini index of the net wealth distribution model

\subsubsection{The Lorenz curve and the Gini index of the net wealth distribution model}
\label{sec:TheLorenzCurveAndTheGiniIndexOfTheNetWealthDistributionModel}

\begin{figure}[!t]
\centering
\subfigure[$m>0$]{\includegraphics[width=0.48\textwidth]{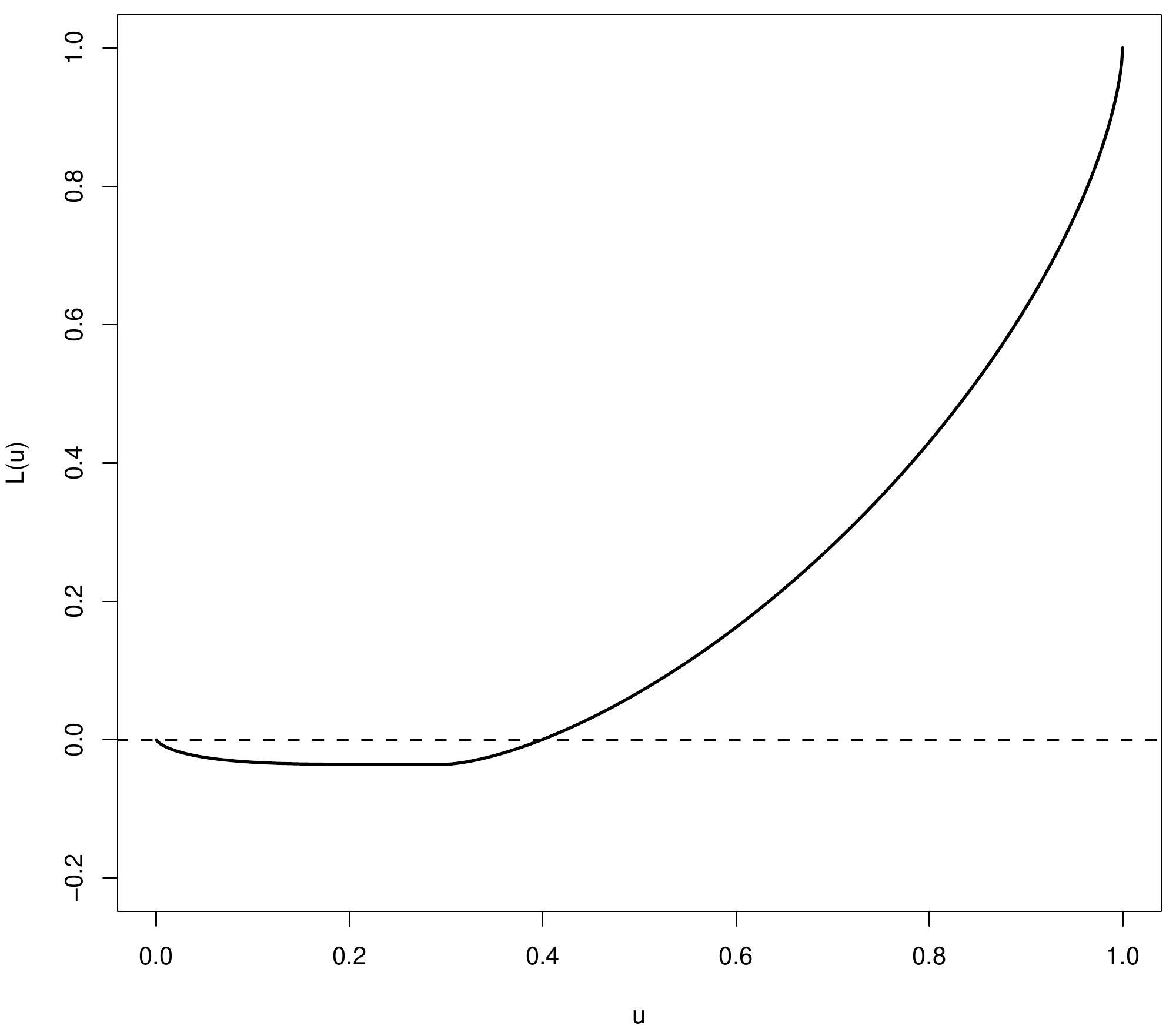}\label{fig:Figure_5_a}}\quad
\subfigure[$m<0$]{\includegraphics[width=0.48\textwidth]{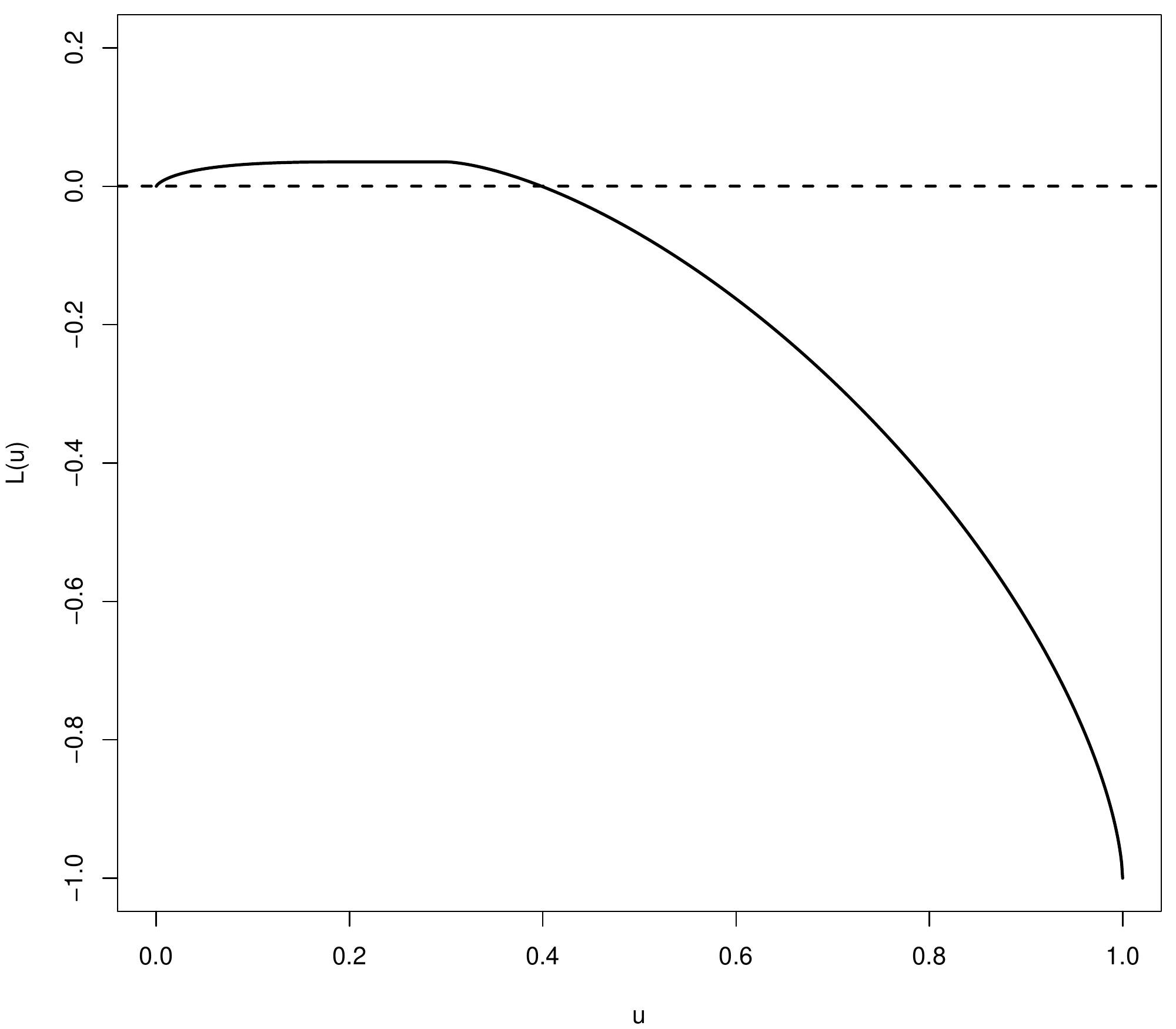}\label{fig:Figure_5_b}}
\caption{The Lorenz curve of the $\kappa$-generalized mixture model for net wealth distribution. Both curves in panels \subref{fig:Figure_5_a} and \subref{fig:Figure_5_b} have been drawn using the following parameter values: $s=0.7$, $\alpha=2$, $\beta=10$, $\kappa=0.75$, $\lambda=1$, $\theta_{1}=0.2$ and $\theta_{2}=0.1$. Mean net wealth $m$ equals, respectively, 7.172 and -7.172.}
\label{fig:Figure_5}
\end{figure}

Given the mathematical structure of the net wealth distribution function specified by \eqref{eq:Equation_36}, we have
\begin{equation}
L\left(u\right)=
\begin{cases}
-\frac{\lambda\theta_{1}}{m}\Gamma\left(1+\frac{1}{s},\log\frac{\theta_{1}}{u}\right)&\text{if $0\leq u<\theta_{1}$},\\
-\frac{\lambda\theta_{1}}{m}\Gamma\left(1+\frac{1}{s}\right)&\text{if $\theta_{1}\leq u\leq\rho$},\\
\begin{aligned}
&\frac{1}{m}\left\{\frac{\left(1-\rho\right)\beta}{\left(2\kappa\right)^{1+\frac{1}{\alpha}}}\left[B\left(\frac{1}{2\kappa}-\frac{1}{2\alpha},1+\frac{1}{\alpha}\right)\right.\right.\\[1ex]
&\left.-B\left(\left[\frac{1-u}{1-\rho}\right]^{2\kappa};\frac{1}{2\kappa}-\frac{1}{2\alpha},1+\frac{1}{\alpha}\right)\right]\\
&\left.-\lambda\theta_{1}\Gamma\left(1+\frac{1}{s}\right)\right\}
\end{aligned}&\text{if $u>\rho$},
\end{cases}
\label{eq:Equation_40}
\end{equation}
with $L\left(1\right)=1$. Equation \eqref{eq:Equation_40} determines the path of the Lorenz curve for the $\kappa$-generalized mixture model over the whole range of wealth. A graphical representation of its shape is given in Figure \ref{fig:Figure_5}.
%
% Figure 5: The Lorenz curve of the k-generalized mixture model for net wealth distribution --------------
% --------------------------------------------------------------------------------------------------------
%
Assuming mean wealth for the whole population is positive, as in panel \subref{fig:Figure_5_a} of the figure, the first part of $L\left(u\right)$ drops below the horizontal line at zero and has a negative slope for negative net wealth values; then the curve is horizontal, corresponding to households that have zero net wealth, and has the conventional positive slope over the remaining units (with positive wealth values). Relative to the conventional picture of a Lorenz curve, the curve \eqref{eq:Equation_40} takes on an even more non-standard shape when $m<0$, as in this case it appears to be flipped vertically\textemdash see panel \subref{fig:Figure_5_b} of the figure. Indeed, starting from the poorest unit, the curve has a positive slope and lies above the zero line over the range of negative wealth values; then it is horizontal where wealth is zero, and has a negative slope over the remaining (positive) wealth units.

Following from the fact that the net wealth Lorenz curve presents negative values for all $u\leq\rho$, it can be proved that the Gini index takes the general form \cite{Dagum2006a,Dagum2006b}
\begin{equation}
G=\frac{2\int\limits^{1}_{0}\left[u-L\left(u\right)\right]\operatorname{d}u}{1+\rho\left|L\left(\theta_{1}\right)\right|}=\frac{1-2\int\limits^{1}_{0}L\left(u\right)\operatorname{d}u}{1-\rho L\left(\theta_{1}\right)},
\end{equation}
where
\begin{equation}
\int\limits^{1}_{0}L\left(u\right)\operatorname{d}u=\int\limits^{\theta_{1}}_{0}L\left(u\right)\operatorname{d}u+\int\limits^{\rho}_{\theta_{1}}L\left(u\right)\operatorname{d}u+\int\limits^{1}_{\rho}L\left(u\right)\operatorname{d}u.
\end{equation}
Using \eqref{eq:Equation_40}, the Gini of the $\kappa$-generalized mixture model for net wealth distribution becomes
\begin{equation}
G=\frac{m-2\left[\frac{\left(1-\rho\right)^{2}\beta}{\left(2\kappa\right)^{1+\frac{1}{\alpha}}}B\left(\frac{1}{\kappa}-\frac{1}{2\alpha},1+\frac{1}{\alpha}\right)-\lambda\theta_{1}\left(1-\theta_{1}2^{-1-\frac{1}{s}}\right)\Gamma\left(1+\frac{1}{s}\right)\right]}{m+\rho\lambda\theta_{1}\Gamma\left(1+\frac{1}{s}\right)}.
\end{equation}
Hence, while $\beta$ and $\lambda$ are scale parameters, all the others\textemdash i.e. $s$, $\alpha$, $\kappa$, $\theta_{1}$, $\theta_{2}$ and $\theta_{3}$\textemdash are inequality parameters. Specifically, $s$, $\theta_{1}$ and $\theta_{2}$ account for the contribution to the Gini index of the negative and null net wealth observations, whereas $\alpha$, $\kappa$ and $\theta_{3}$ account for the contribution of the positive observations.

The Gini coefficient of net wealth distribution is well-defined when households have negative or zero wealth holdings, but estimates of the coefficient might be greater than one if those households represent a relatively large fraction of the population; furthermore, whenever the mean net wealth is negative, estimates of the index are even negative \cite{Hagerbaumer1977,PyattChenFei1980,AmielCowellPolovin1996,JenkinsJantti2005,Cowell2011}.

% Four-parameter extensions of the κ-generalized distribution --------------------------------------------

\subsection{Four-parameter extensions of the $\boldsymbol{\kappa}$-generalized distribution}
\label{sec:FourParameterExtensionsOfTheKappaGeneralizedDistribution}

Quite recently, two four-parameter extensions of the $\kappa$-generalized distribution were introduced by \cite{Okamoto2013} and named \textit{extended $\kappa$-generalized distributions of the first and second kind} (or E$\kappa$G1 and E$\kappa$G2, for short).

The first kind of generalization, the E$\kappa$G1, was obtained by ``Weibullizing'' a two-parameter deformed exponential, implicitly defined as the inverse of the following two-parameter generalization of the logarithm function \cite{KaniadakisLissiaScarfone2004,KaniadakisLissiaScarfone2005}
\begin{equation}
\ln_{\left\{\kappa,r\right\}}\left(x\right)=x^{r}\frac{x^{\kappa}-x^{-\kappa}}{2\kappa},
\end{equation}
which recovers the standard logarithm in the limit $\left(\kappa,r\right)\rightarrow\left(0,0\right)$ independently of the direction.

No closed-form expression of the cumulative distribution function is available for the E$\kappa$G1, while the quantile function is given by the pleasantly simple formula
\begin{equation}
F^{-1}\left(u;a,b,q,r\right)=b\left[-\left(1-u\right)^{r}\frac{\left(1-u\right)^{\frac{1}{2q}}-\left(1-u\right)^{-\frac{1}{2q}}}{1/q}\right]^{\frac{1}{a}},\;\;\, 0<u<1,
\label{eq:Equation_45}
\end{equation}
where $a,b,q>0$ and $r<\frac{1}{2q}$. When $r=0$ (and $a=\alpha$, $b=\beta$, $q=\frac{1}{2\kappa}$) the E$\kappa$G1 quantile function is equivalent to that of the $\kappa$-generalized distribution\textemdash compare with \eqref{eq:Equation_14}.

The probability density function of the E$\kappa$G1 can be expressed in terms of the cumulative probabilities $u$ as follows\footnote{A probability density function expressed in terms of $u$ was called the ``density quantile function'' by \cite{Parzen1979}.}
\begin{equation}
f_{u}\left(u;a,b,q,r\right)=\frac{a\left[-\left(1-u\right)^{r}\frac{\left(1-u\right)^{\frac{1}{2q}}-\left(1-u\right)^{-\frac{1}{2q}}}{1/q}\right]^{-\frac{1}{a}+1}}{b\left[\left(qr+\frac{1}{2}\right)\left(1-u\right)^{r+\frac{1}{2q}-1}-\left(qr-\frac{1}{2}\right)\left(1-u\right)^{r-\frac{1}{2q}-1}\right]}.
\label{eq:Equation_46}
\end{equation}
Since $x=F^{-1}\left(u;a,b,q,r\right)$ and $u=F\left(x;a,b,q,r\right)$ for any pair of values $\left(x,u\right)$, it follows from the definition of differentiation that
\begin{equation}
f_{u}\left(u;a,b,q,r\right)=\frac{1}{\frac{\operatorname{d}F^{-1}\left(u;a,b,q,r\right)}{\operatorname{d}u}}=\frac{\operatorname{d}F\left(x;a,b,q,r\right)}{\operatorname{d}x}=f\left(x;a,b,q,r\right).
\label{eq:Equation_47}
\end{equation}
\sloppy This result provides a way of plotting the density function \eqref{eq:Equation_46}. Indeed, if we let $u$ take the values, say, $u=0.01,0.02,\ldots,0.99$, and plot the points $\left(F^{-1}\left(u;a,b,q,r\right),f_{u}\left(u;a,b,q,r\right)\right)$, then we get the plot of points $\left(x,f\left(x;a,b,q,r\right)\right)$, i.e. the plot of the density of $x$ given by \eqref{eq:Equation_47}. Thus plots of the E$\kappa$G1 probability density function can be obtained entirely from the quantile function \eqref{eq:Equation_45}.

The E$\kappa$G1 density follows a power-law behavior in both of its tails, i.e.
\begin{equation}
f\left(x;a,b,q,r\right)\underset{x\to0}{\sim}c_{1}x^{a-1}\;\;\,\text{and}\;\;\,f\left(x;a,b,q,r\right)\underset{x\to+\infty}{\sim}c_{2}x^{-\frac{a}{\left(\frac{1}{2q}-r\right)-1}},
\label{eq:Equation_5_5}
\end{equation}
where $c_{1}$ and $c_{2}$ are positive constants. The upper-tail power-law behavior is simply the Pareto law, whereas the power-law behavior in the lower tail of the distribution is a stylized fact that sometimes seems to be born out in actual data \cite{Reed2003,Reed2004} and was indeed identified many years ago by \cite{Champernowne1953}.

\begin{figure}[!p]
\centering
\subfigure[$a=3.00$, $b=1.00$, $q=0.60$ and $r=0.30$]{\includegraphics[width=0.48\textwidth]{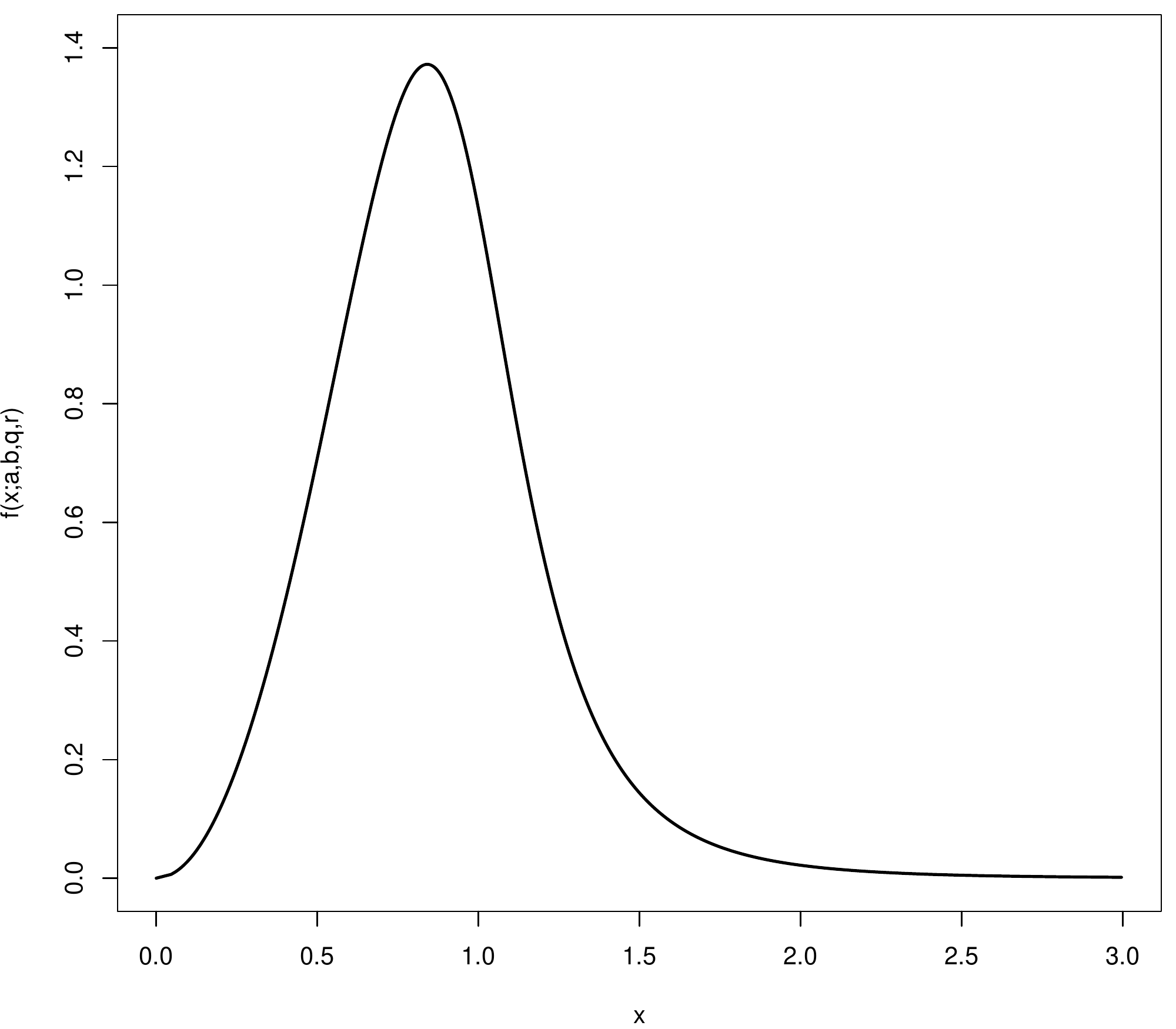}\label{fig:Figure_6_a}}\quad
\subfigure[$a=3.00$, $b=1.00$, $q=0.60$ and $r=0.80$]{\includegraphics[width=0.48\textwidth]{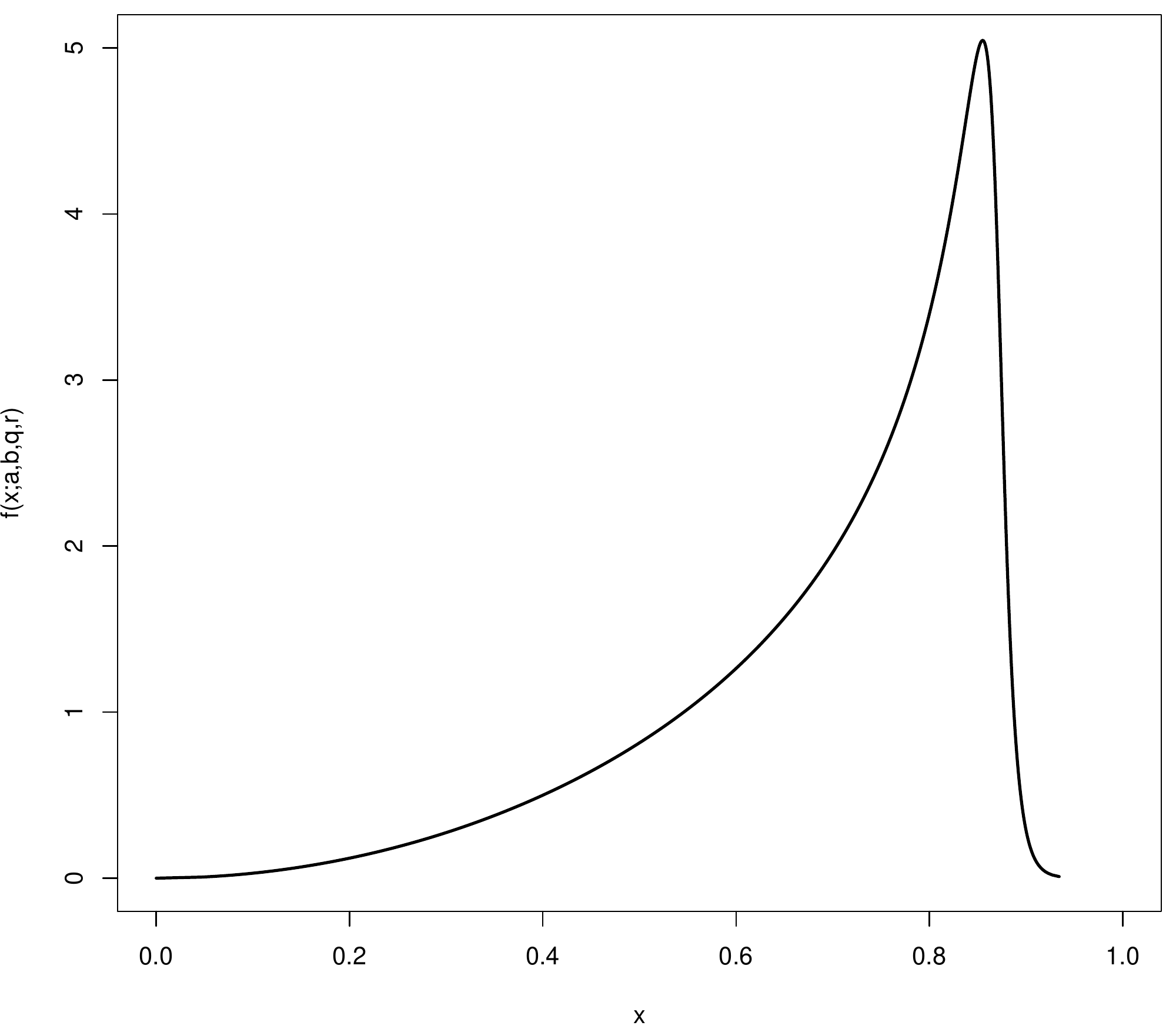}\label{fig:Figure_6_b}}\\
\subfigure[$a=1.00$, $b=1.00$, $q=0.60$ and $r=0.30$]{\includegraphics[width=0.48\textwidth]{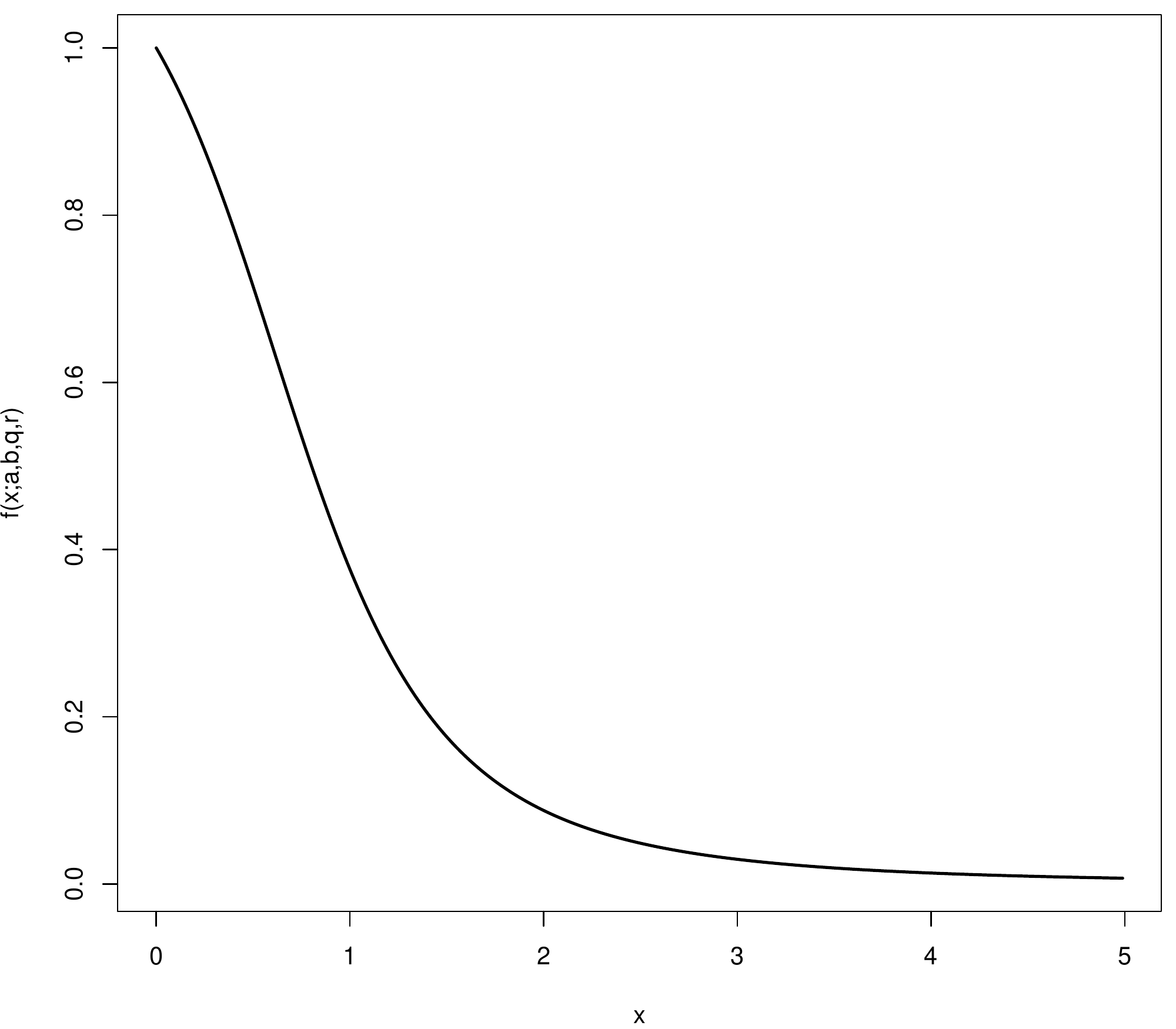}\label{fig:Figure_6_c}}\quad
\subfigure[$a=1.00$, $b=1.00$, $q=0.60$ and $r=0.80$]{\includegraphics[width=0.48\textwidth]{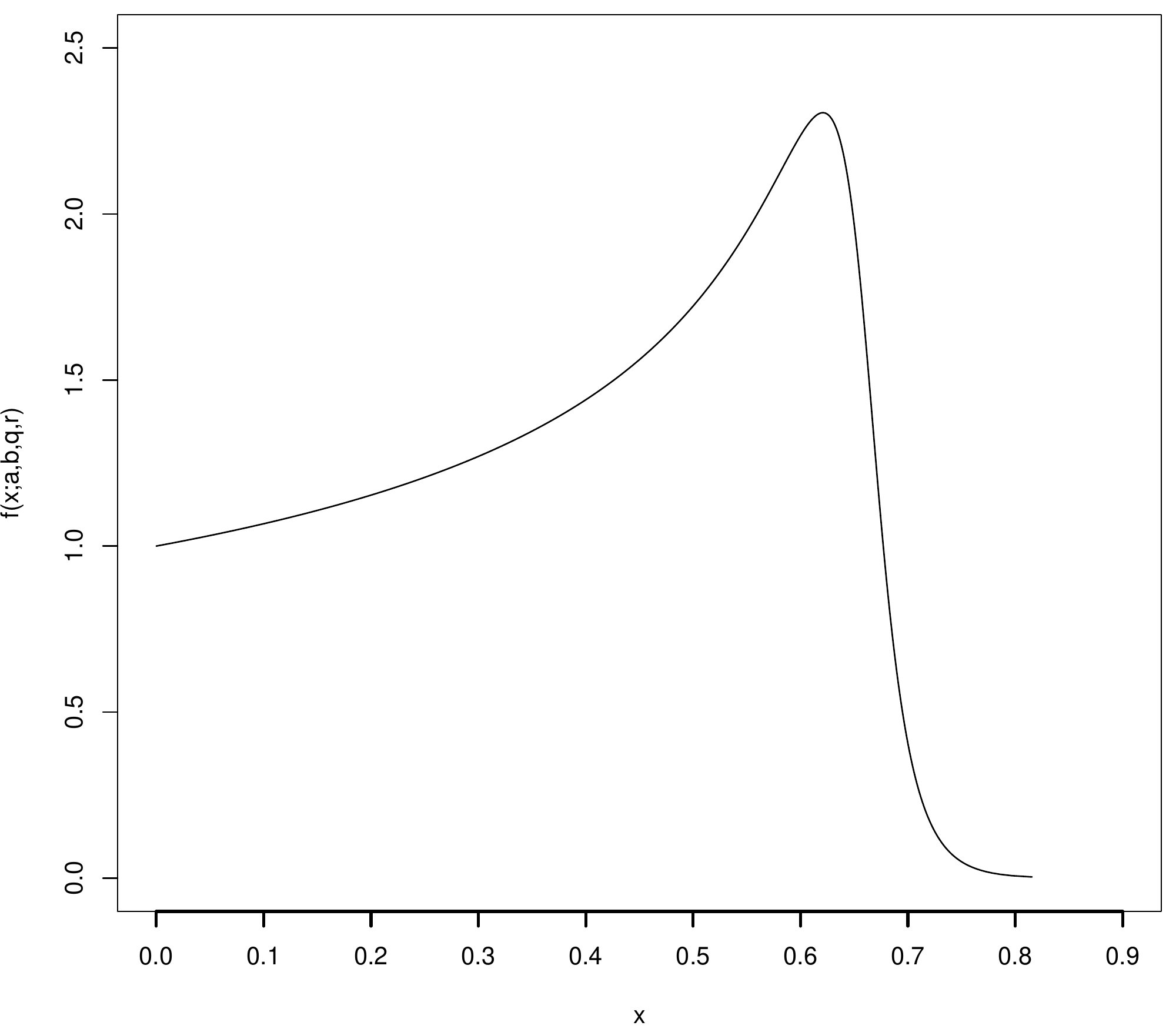}\label{fig:Figure_6_d}}\\
\subfigure[$a=0.80$, $b=1.00$, $q=0.60$ and $r=0.30$]{\includegraphics[width=0.48\textwidth]{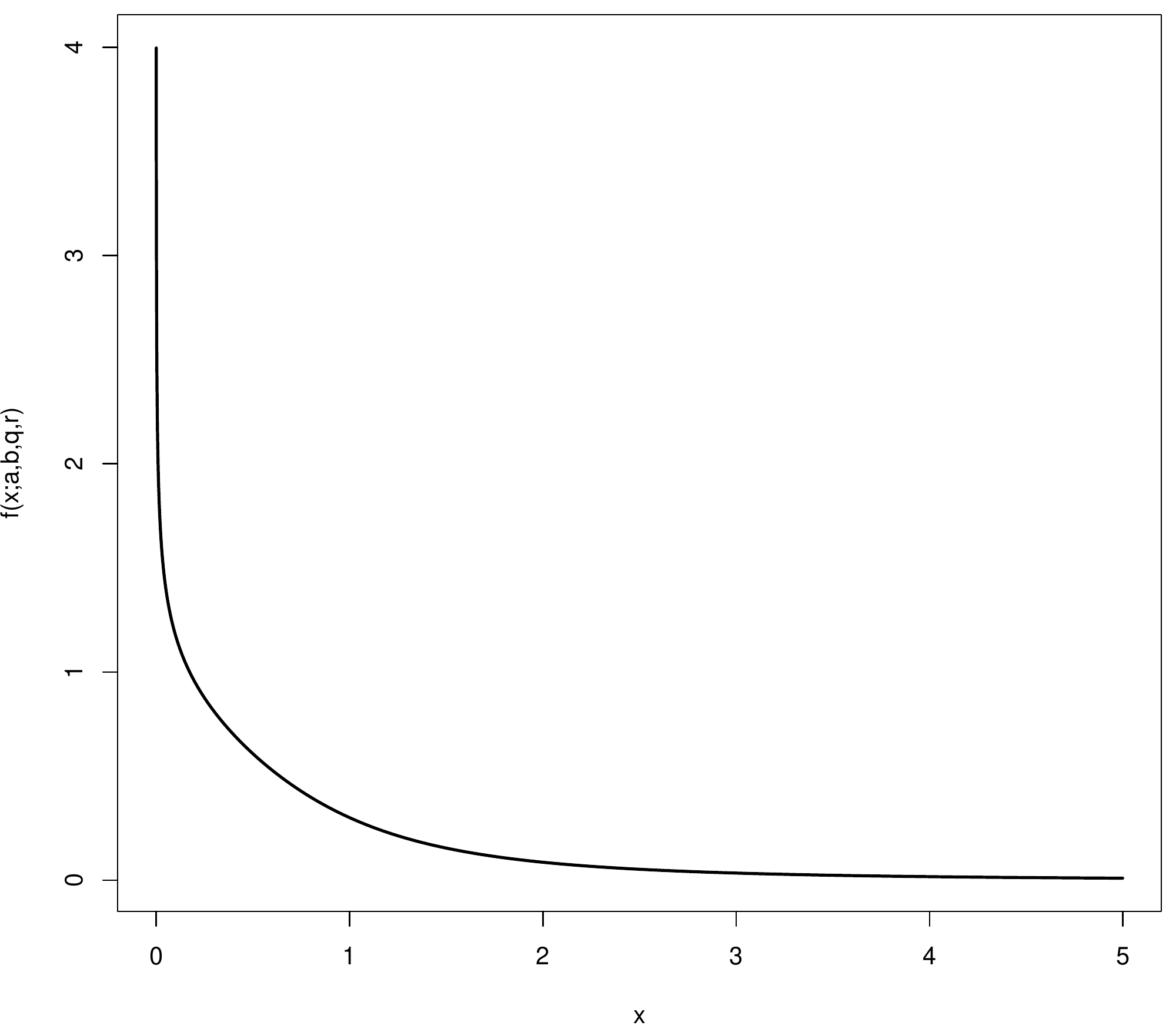}\label{fig:Figure_6_e}}\quad
\subfigure[$a=0.80$, $b=1.00$, $q=0.60$ and $r=0.80$]{\includegraphics[width=0.48\textwidth]{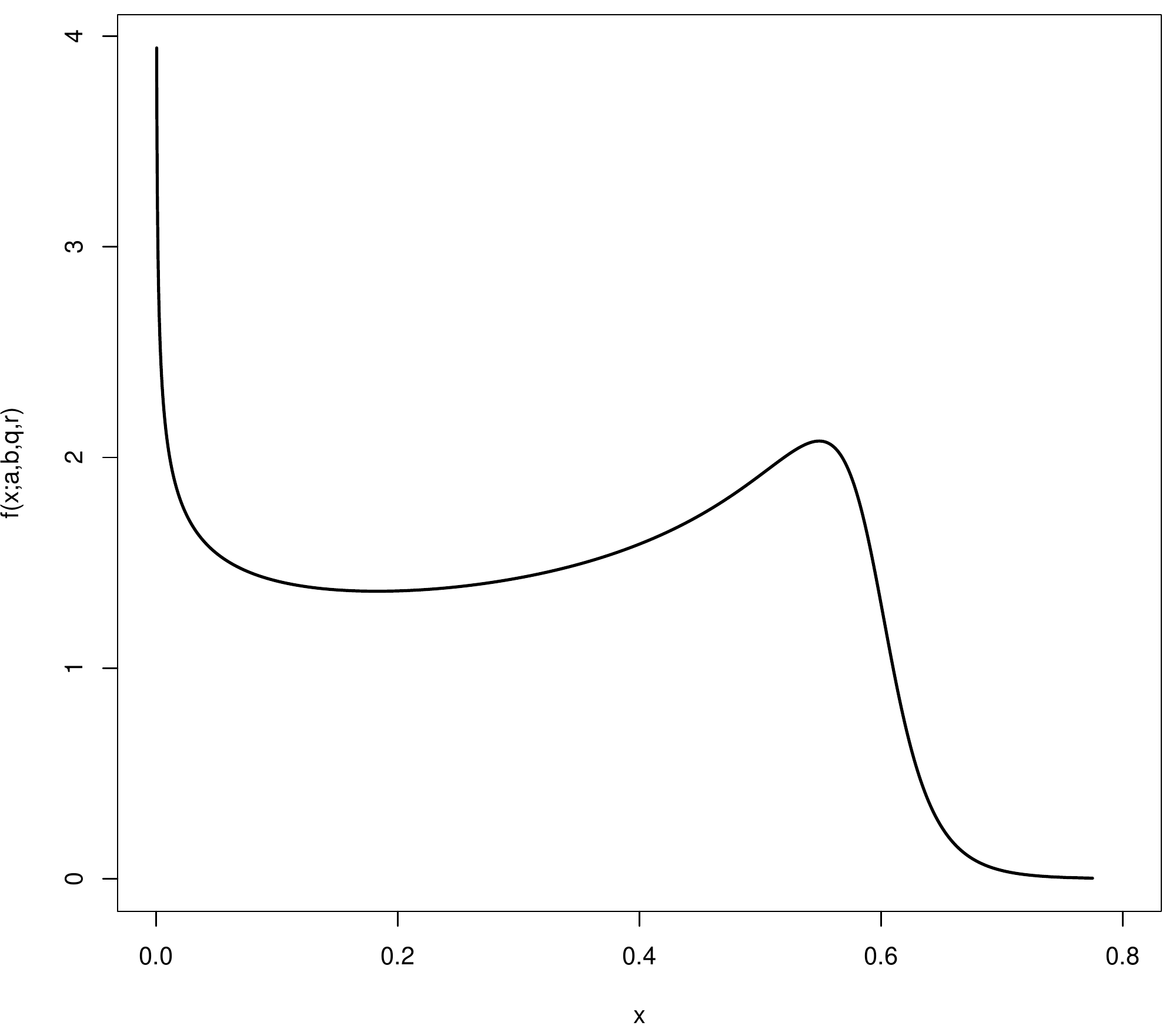}\label{fig:Figure_6_f}}
\caption{E$\kappa$G1 density for various parameter values.}
\label{fig:Figure_6}
\end{figure}

Figure \ref{fig:Figure_6} illustrates the shape of the E$\kappa$G1 density for various parameter values.
%
% Figure 6: EκG1 density for various parameter values ----------------------------------------------------
% --------------------------------------------------------------------------------------------------------
%
In the case $a>1$, as in panels \subref{fig:Figure_6_a} and \subref{fig:Figure_6_b}, $f\left(x;a,b,q,r\right)\to0$ when $x\to0$ and there is always an interior mode. By contrast, if $a=1$, $f\left(x;a,b,q,r\right)\to\frac{a}{b}$ when $x\to0$, i.e. the E$\kappa$G1 density approaches a finite positive value at the left limit; furthermore, the density is monotonically decreasing for $r\leq\frac{1}{2}$, whereas for $r>\frac{1}{2}$ it permits an interior mode\textemdash see panels \subref{fig:Figure_6_c} and \subref{fig:Figure_6_d}, respectively. Finally, for $a<1$ the E$\kappa$G1 density is infinite at the left limit, i.e. $f\left(x;a,b,q,r\right)\to+\infty$ when $x\to0$, and it can be either monotonically decreasing\textemdash as in panel \subref{fig:Figure_6_e}\textemdash or have a local maximum and minimum\textemdash as in panel \subref{fig:Figure_6_f}.

The second generalization, the E$\kappa$G2, was derived by extending the $\kappa$-generalized Lorenz curve
\begin{equation}
L\left(u\right)=I_{z}\left(1+\frac{1}{\alpha},\frac{1}{2\kappa}-\frac{1}{2\alpha}\right),\;\;\,z=1-\left(1-u\right)^{2\kappa}=I^{-1}_{u}\left(1,\frac{1}{2\kappa}\right)
\end{equation}
in the following manner
\begin{equation}
L\left(u\right)=I_{z}\left(p+\frac{1}{a},q-\frac{1}{2a}\right),\;\;\,z=I^{-1}_{u}\left(p,q\right),
\end{equation}
where $a=\alpha$, $q=\frac{1}{2\kappa}$ and $I^{-1}_{u}\left(\cdot,\cdot\right)$ denotes the inverse of the regularized incomplete beta function.

The cumulative distribution function of the E$\kappa$G2 is available in closed form. It can be expressed in terms of the incomplete beta function ratio as follows
\begin{equation}
F\left(x;a,b,p,q\right)=I_{z}\left(p,q\right),\;\;\,z=\left(\frac{x}{b}\right)^{a}\left[\sqrt{1+\frac{1}{4}\left(\frac{x}{b}\right)^{2a}}-\frac{1}{2}\left(\frac{x}{b}\right)^{a}\right],\;\;\,x>0,
\label{eq:Equation_51}
\end{equation}
where all four parameters $a$, $b$, $p$ and $q$ are positive. Equation \eqref{eq:Equation_51} implies that even the quantile function is available in closed form; it is
\begin{equation}
F^{-1}\left(x;a,b,p,q\right)=bz^{\frac{1}{a}}\left(1-z\right)^{-\frac{1}{2a}},\;\;\,z=I^{-1}_{u}\left(p,q\right).
\label{eq:Equation_52}
\end{equation}
For $p=1$ (and $a=\alpha$, $b=\beta$, $q=\frac{1}{2\kappa}$), it follows from \eqref{eq:Equation_51} and \eqref{eq:Equation_52} that the E$\kappa$G2 is equivalent to the $\kappa$-generalized distribution.

The probability density function of the E$\kappa$G2 has the form
\begin{equation}
f\left(x;a,b,p,q\right)=\frac{a}{bB\left(p,q\right)}\frac{z^{p-\frac{1}{a}}\left(1-z\right)^{q+\frac{1}{2a}}}{1-\frac{1}{2}z},
\label{eq:Equation_53}
\end{equation}
where $z=\left(\frac{x}{b}\right)^{a}\left[\sqrt{1+\frac{1}{4}\left(\frac{x}{b}\right)^{2a}}-\frac{1}{2}\left(\frac{x}{b}\right)^{a}\right]$. The lower and upper tails of \eqref{eq:Equation_53} exhibit power-law (Paretian) behavior, i.e.
\begin{equation}
f\left(x;a,b,p,q\right)\underset{x\to0}{\sim}c_{3}x^{ap-1}\;\;\,\text{and}\;\;\,f\left(x;a,b,p,q\right)\underset{x\to+\infty}{\sim}c_{4}x^{-2aq-1}
\end{equation}
for constants $c_{3}$ and $c_{4}$.

\begin{figure}[!p]
\centering
\subfigure[$a=2.00$, $b=1.00$, $p=1.00$ and $q=1.00$]{\includegraphics[width=0.48\textwidth]{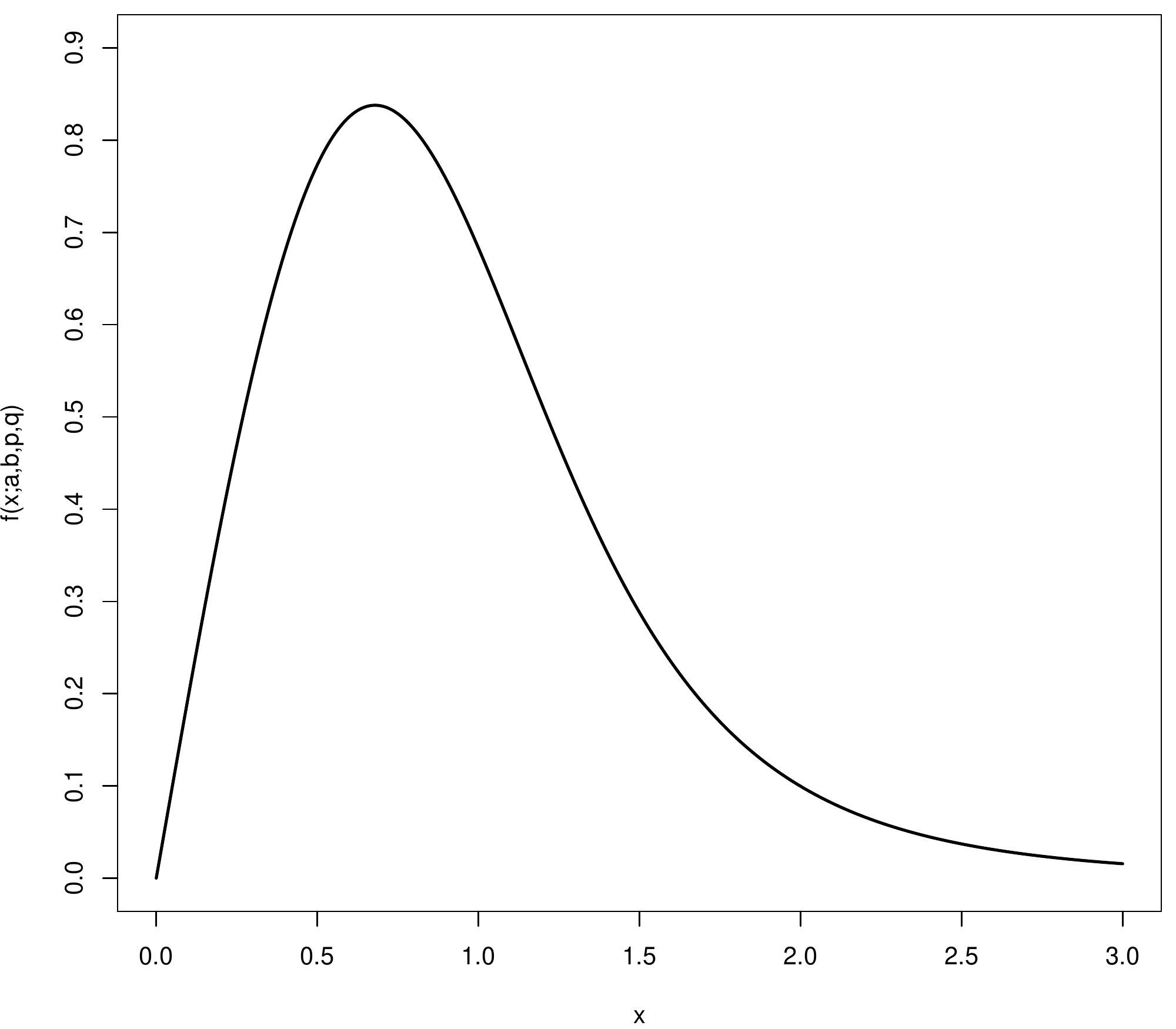}\label{fig:Figure_7_a}}\quad
\subfigure[$a=5.00$, $b=1.00$, $p=1.00$ and $q=1.00$]{\includegraphics[width=0.48\textwidth]{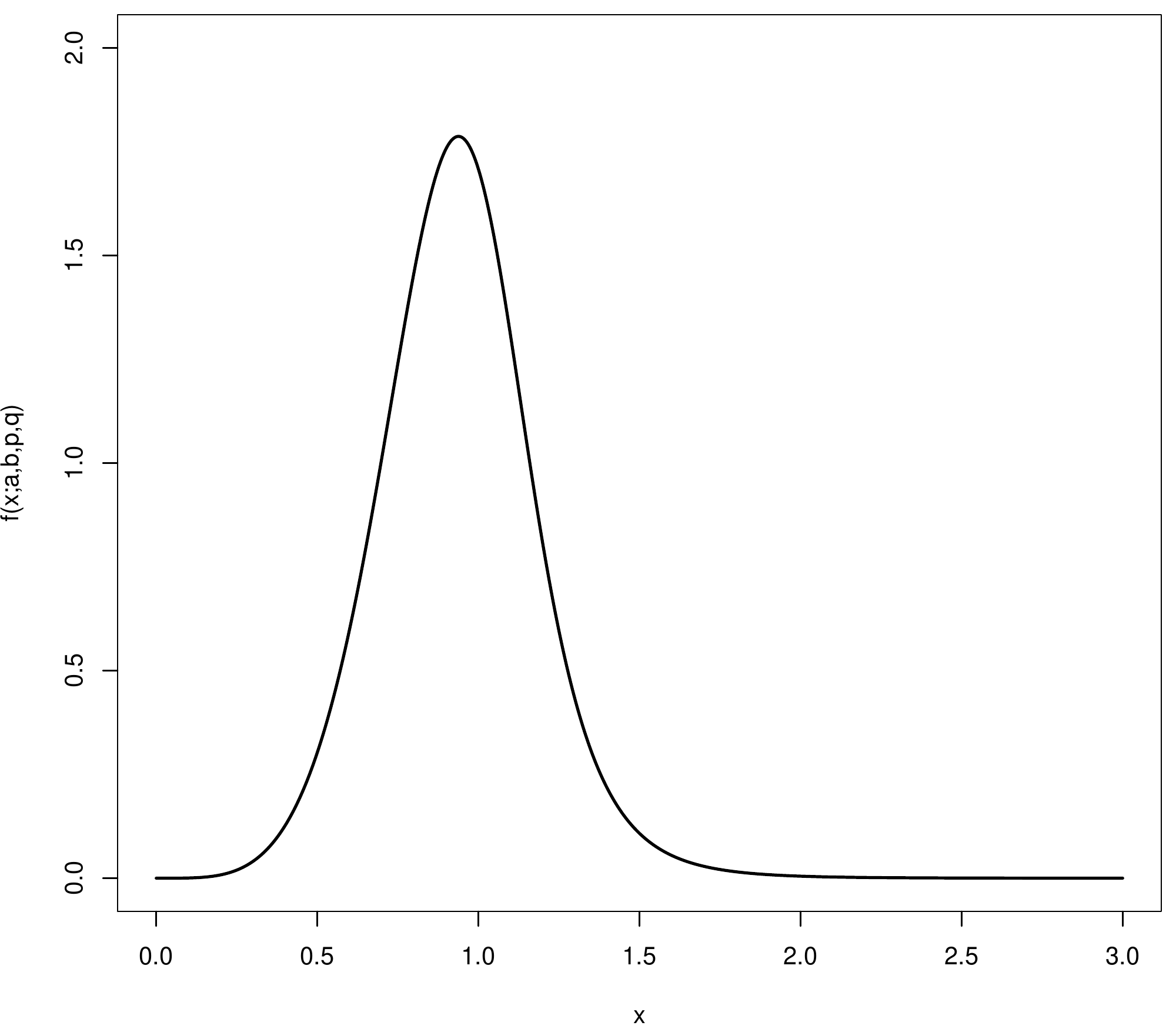}\label{fig:Figure_7_b}}\\
\subfigure[$a=2.00$, $b=1.00$, $p=0.50$ and $q=0.25$]{\includegraphics[width=0.48\textwidth]{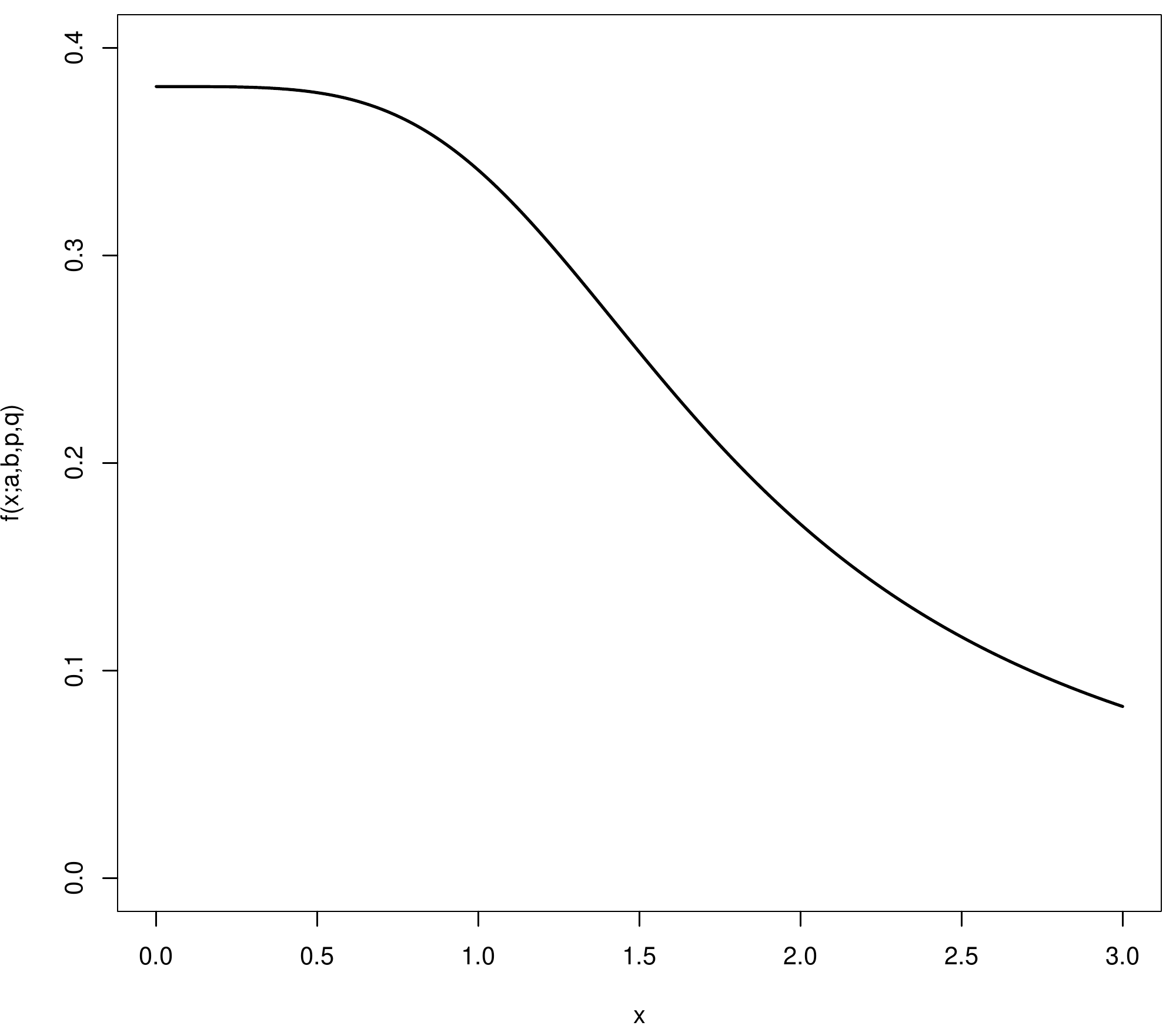}\label{fig:Figure_7_c}}\quad
\subfigure[$a=5.00$, $b=1.00$, $p=0.20$ and $q=0.25$]{\includegraphics[width=0.48\textwidth]{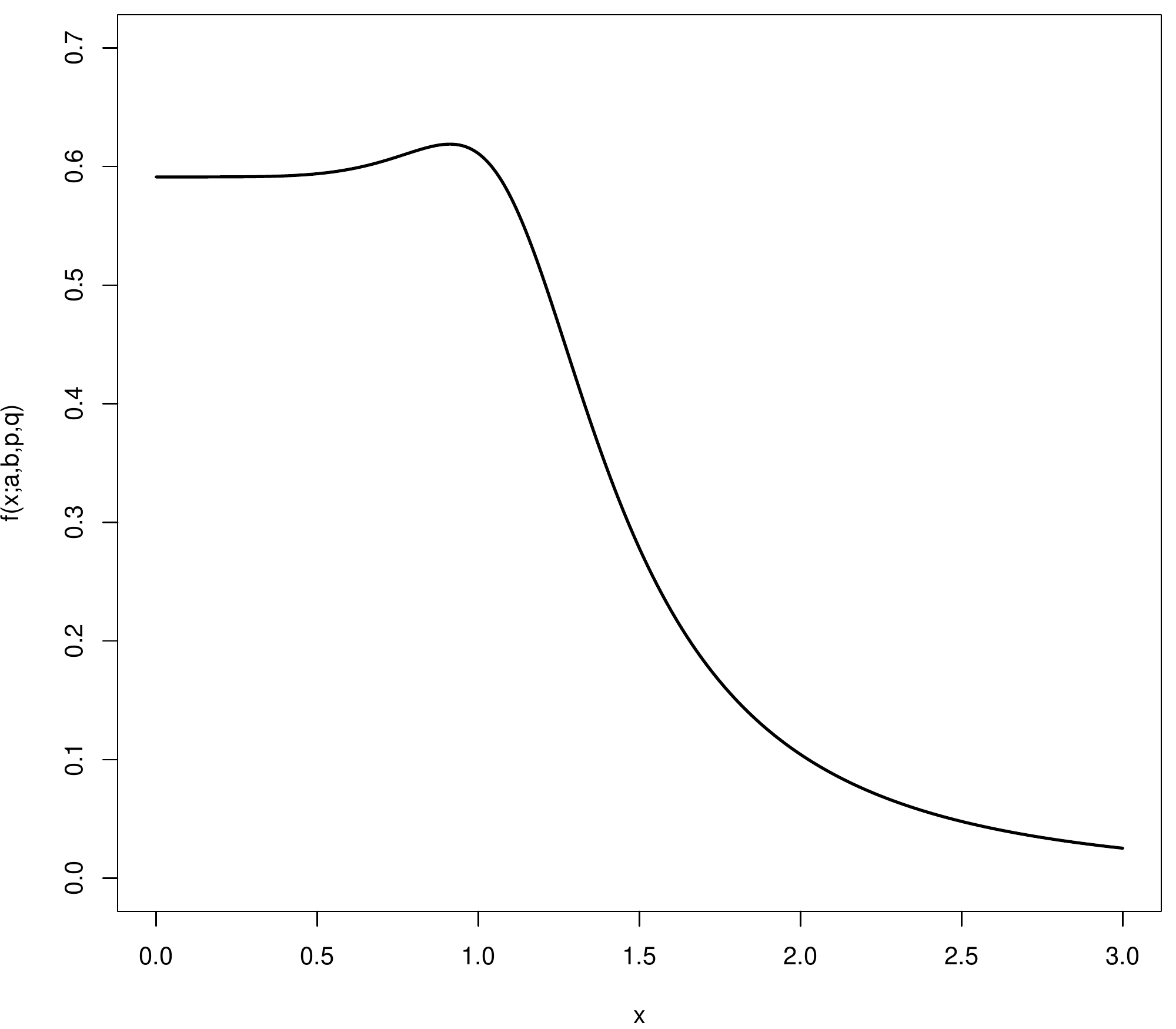}\label{fig:Figure_7_d}}\\
\subfigure[$a=2.00$, $b=1.00$, $p=0.10$ and $q=0.10$]{\includegraphics[width=0.48\textwidth]{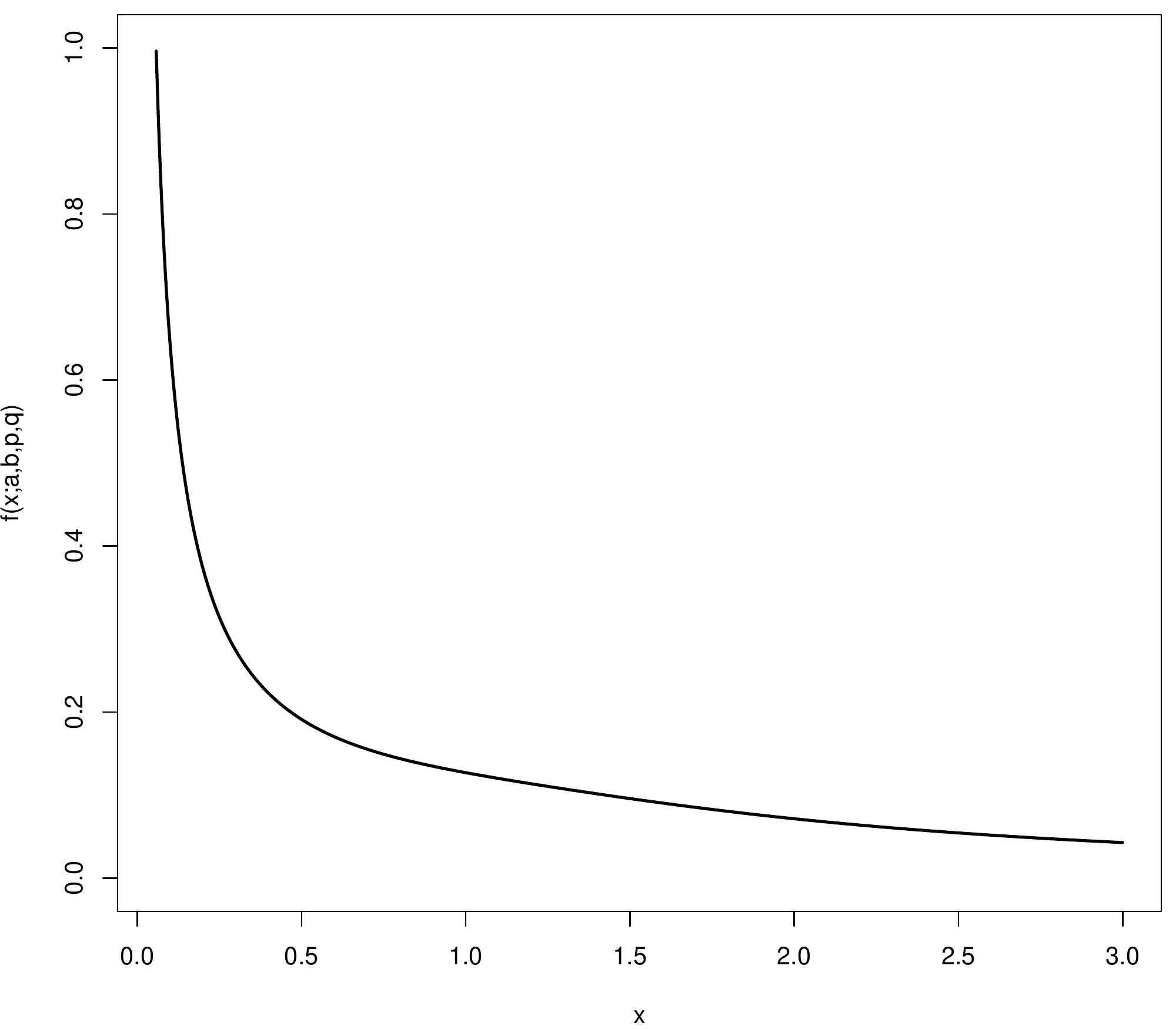}\label{fig:Figure_7_e}}\quad
\subfigure[$a=5.00$, $b=1.00$, $p=0.10$ and $q=0.10$]{\includegraphics[width=0.48\textwidth]{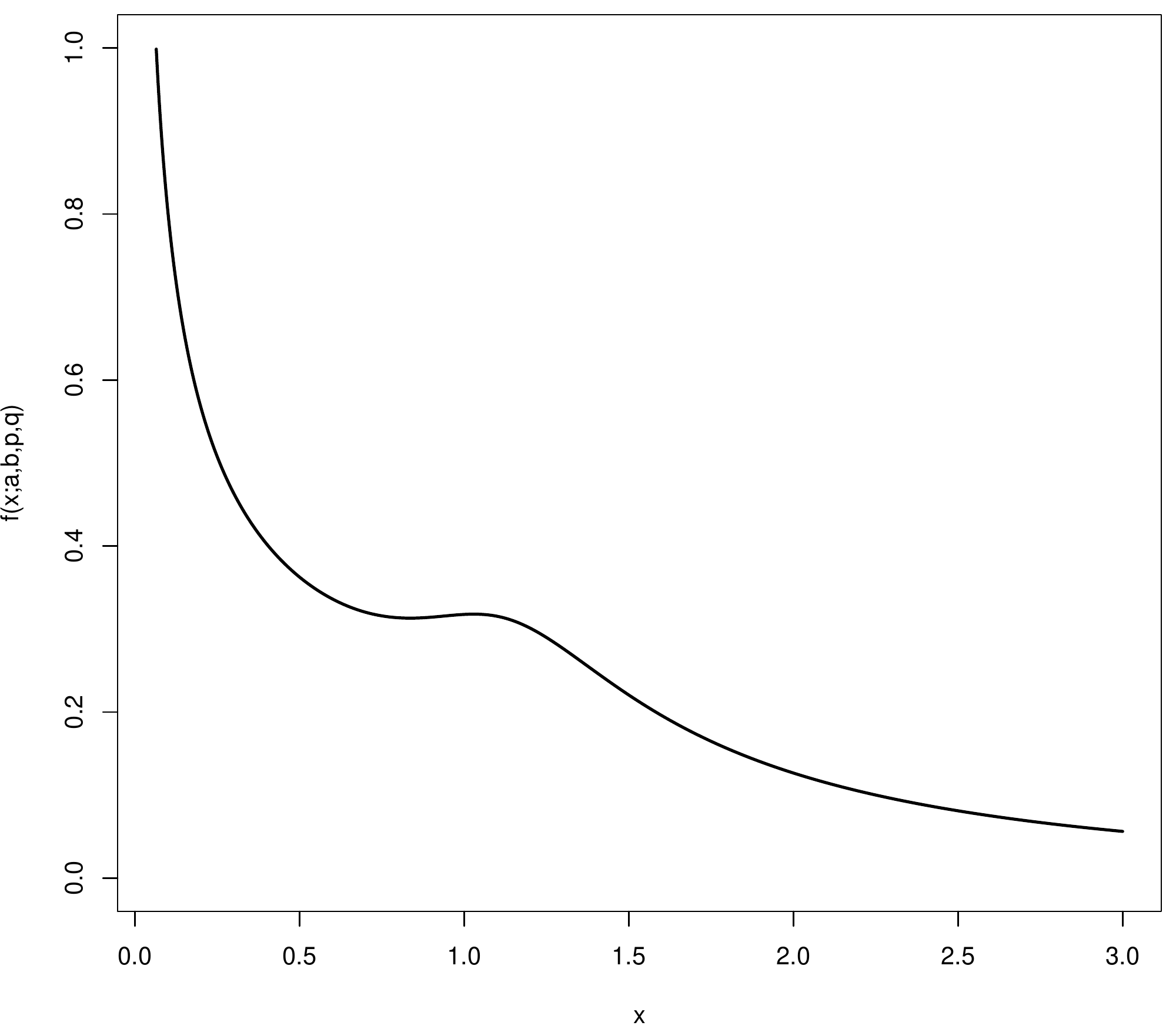}\label{fig:Figure_7_f}}
\caption{E$\kappa$G2 density for various parameter values.}
\label{fig:Figure_7}
\end{figure}

Figure \ref{fig:Figure_7} charts the E$\kappa$G2 density for various parameter values.
%
% Figure 7: EκG2 density for various parameter values ----------------------------------------------------
% --------------------------------------------------------------------------------------------------------
%
In the case $p>\frac{1}{a}$, as in panels \subref{fig:Figure_7_a} and \subref{fig:Figure_7_b}, $f\left(x;a,b,p,q\right)\to0$ when $x\to0$ and the density always exhibits an interior mode. By contrast, if $p=\frac{1}{a}$, $f\left(x;a,b,p,q\right)\to\frac{a}{bB\left(p,q\right)}$ when $x\to0$, i.e. the E$\kappa$G2 density approaches a finite positive value at the left limit; furthermore, the density is monotonically decreasing for $q+\frac{1}{2a}\geq\frac{1}{2}$, whereas for $q+\frac{1}{2a}<\frac{1}{2}$ it permits an interior mode\textemdash see panels \subref{fig:Figure_7_c} and \subref{fig:Figure_7_d}, respectively. Finally, for $p<\frac{1}{a}$ the E$\kappa$G2 density is infinite at the left limit, i.e. $f\left(x;a,b,p,q\right)\to+\infty$ when $x\to0$, and it can be either monotonically decreasing\textemdash as in panel \subref{fig:Figure_7_e}\textemdash or have a local maximum and minimum\textemdash as in panel \subref{fig:Figure_7_f}.

Statistical tools for describing distributions and analyzing inequality can be expressed analytically for the four-parameter variants of the $\kappa$-generalized distribution. References \cite{Okamoto2013,ClementiGallegati2016} reports formulas for the moments, the Lorenz curve, the Gini index, the coefficient of variation, the mean logarithmic deviation and the Theil index for both of the models presented. In addition, they also discuss distributional properties for the reciprocal of a random variable from the E$\kappa$G1 or E$\kappa$G2 distribution.

% APPLICATIONS OF κ-GENERALIZED MODELS TO INCOME AND WEALTH DATA -----------------------------------------

\section{\texorpdfstring{Applications of $\boldsymbol{\kappa}$-generalized models to income and wealth data}{Applications of κ-generalized models to income and wealth data}}
\label{sec:ApplicationsOfKappaGeneralizedModelsToIncomeAndWealthData}

During the last decade, there have been several applications of $\kappa$-generalized models to real-world data on income and wealth distribution.

The earliest investigation was completed by \cite{ClementiGallegatiKaniadakis2007}, who applied re-parametrized versions of \eqref{eq:Equation_1} and \eqref{eq:Equation_2} to 2001--02 household incomes for Germany, Italy, and the United Kingdom. They found an excellent agreement between the model and the whole spectrum of the empirical distributions, from the low to the high incomes, including the intermediate region for which a clear deviation was discovered when the two liming cases of the Weibull model \eqref{eq:Equation_11} and the pure Pareto power-law \eqref{eq:Equation_12} were used for the same interpolation purposes.

Subsequently, the same parametrization of the $\kappa$-generalized distribution was fitted to Australian household incomes for 2002--03 \cite{ClementiDiMatteoGallegatiKaniadakis2008} and the US family incomes for 2003 \cite{ClementiDiMatteoGallegatiKaniadakis2008,ClementiGallegatiKaniadakis2009}. Once again, the model was found to describe the entire income range extremely well; in addition, it was discovered to provide very accurate estimates of the inequality level for both the countries as evaluated by means of the Lorenz curve and the associated Gini measure of inequality.

Of special interest are papers fitting several distributions to the same data, with an eye on relative performance. From comparative studies such as \cite{ClementiGallegatiKaniadakis2010}, who considered the distribution of household income in Italy for the years 1989 to 2006, it emerges that model \eqref{eq:Equation_2} typically outperforms its three-parameter competitors such as the Singh-Maddala \cite{SinghMaddala1976} and Dagum type I \cite{Dagum1977} distributions, apart from the GB2 which has an extra parameter.\footnote{The GB2 is a quite general family of parametric models for the size distribution of income introduced by \cite{McDonald1984} that nests most of the functional forms previously considered in the size distributions literature as special or limiting cases. In particular, both the Singh-Maddala and Dagum type I distributions are special cases of the GB2.} The model was also fitted by \cite{ClementiGallegatiKaniadakis2012a} to data from other household budget surveys, namely Germany 1984--2007, Great Britain 1991--2004, and the United States 1980--2005. In a remarkable number of cases, the distribution of household income follows the $\kappa$-generalized more closely than the Singh-Maddala and Dagum type I. In particular, the fit is statistically superior in the right tail of data with respect to the other competitors in many instances. Another example of comparative study is \cite{Okamoto2012}, who considered US and Italian income data for the 2000s. He found the three-parameter $\kappa$-generalized model to yield better estimates of income inequality even when the goodness-of-fit is inferior to that of distributions in the GB2 family. The excellent fit of the $\kappa$-generalized distribution and its ability in providing relatively more accurate estimation of income inequality were recently confirmed in a book by \cite{ClementiGallegati2016}, who utilize household income data for 45 countries selected from the most recent waves of the LIS Database (\url{http://www.lisdatacenter.org/}).

The previously mentioned works were mainly concerned with the distribution of households incomes. In an interesting contribution by \cite{ClementiGallegatiKaniadakis2012b}, the $\kappa$-generalized distribution was used in a three-component mixture to model the US net wealth data for 1984--2011. Both graphical procedures and statistical methods indicate an overall good approximation of the data. The authors also highlight the relative merits of their specification with respect to finite mixture models based upon the Singh-Maddala and Dagum type I distributions for the positive values of net wealth. Similar results were recently obtained by \cite{ClementiGallegati2016} when analyzing net wealth data for 9 countries selected from the most recent waves of the LWS Database (\url{http://www.lisdatacenter.org/}).

Finally, the four-parameter models discussed in Section \ref{sec:FourParameterExtensionsOfTheKappaGeneralizedDistribution} were used by \cite{Okamoto2013} to analyze household income/consumption data for approximately 20 countries selected from Waves IV to VI of the LIS Database. The data were grouped into 22 classes and 6 different welfare variables were considered for each country dataset. To provide a comparison with alternative four-parameter models of income distribution, the GB2 and the double Pareto-lognormal (dPlN) distribution introduced by \cite{ReedJorgensen2004} were also fitted to the same datasets and welfare variables. In all cases, parameter estimates were obtained using the maximum likelihood method for grouped data and the goodness of fit assessed using both frequency-based (FB) evaluation criteria\textemdash such as the log-likelihood value\textemdash and money-amount-based (MAB) measures, which are expected to more closely reflect the accuracy of inequality estimates\textemdash some examples of MAB measures are the square root of the sum of squared errors between the observed and estimated Lorenz curve (LRSSE) and the absolute error between the observed and estimated Gini index (AEG).

The E$\kappa$G1 is inferior to the E$\kappa$G2, GB2 and dPlN in terms of the log-likelihood value, but superior to the E$\kappa$G2 and GB2 according to the LRSSE and AEG in the overall evaluation. As for the comparison between the E$\kappa$G1 and dPlN in terms of MAB criteria, the former appears to be outperformed by the latter in terms of both the LRSSE and AEG.

No matter how the goodness-of-fit measures are combined, the pairwise comparisons between the E$\kappa$G2 and GB2 show that the former clearly outperforms the latter in the vast majority of cases. The E$\kappa$G2 is also dominant over the dPlN in terms of the log-likelihood value and the LRSSE for both the income and consumption variables, whereas the AEG tends to slightly favor the dPlN.

On the whole, the E$\kappa$G2 outperforms other four-parameter income distribution models. In particular, the E$\kappa$G2 is dominant over its counterparts in almost all cases, although slightly inferior to the dPlN when considering the accuracy of the estimated Gini index. Since the E$\kappa$G2 is close in form to the GB2, which is widely recognized as providing an excellent description of income distributions, this four-parameter variant of the $\kappa$-generalized distribution appears to drill a significant extension to the utility of parametric models for analysis of income distributions.

% DEFORM TO GO STRANGE: AN EXPLANATION OF WHY κ-GENERALIZED MODELS ARE A GOOD FIT TO INCOME AND WEALTH ---
% DISTRIBUTIONS ------------------------------------------------------------------------------------------

\section{\texorpdfstring{Deform to go \textit{strange}: an explanation of why $\boldsymbol{\kappa}$-generalized models are a good fit to income and wealth distributions}{Deform to go strange: an explanation of why κ-generalized models are a good fit to income and wealth distributions}}
\label{sec:DeformToGoStrangeAnExplanationOfWhyKappaGeneralizedModelsAreAGoodFitToIncomeAndWealthDistributions}

As discussed in previous section, parametric models belonging to the $\kappa$-generalized family provide a very good description over the entire income/wealth range, including the upper tail, and the inequality analysis expressed in terms of their parameters reveal very powerful. Lots of parametric models have been formalized the field of income and wealth distribution; the family of $\kappa$-generalized models, we think, is the most appropriate because it is very comfortable with the description of ``strange'' quantities such as income and wealth \cite{Landini2016}.

Intuitively, and generally enough, a quantity is ``strange'' if its probability field is not homogeneously dominated by the same statistical principle over the whole support. That is: depending on the orders of magnitude of the measurement-events, different parts of the realization field are domain of different statistical principles. Stated otherwise, a quantity is strange if below some given order of magnitude no \textit{exclusion principle} is operating, while it is so above. Real sample data show that there is more difference between the richest and the second richest receivers than the poorest and the second poorest, thus suggesting that income and wealth are strange quantities: this is due to the fact that earning and saving capabilities are relative to the income and wealth order of magnitude. Quantities which are dominated by the same statistical principle independently of the part of the income/wealth support the analysis is focused on are said to be \textit{regular}, whereas those dominated by different statistical principles depending on the orders of magnitude are instead said to be \textit{strange}.

By considering the underlying random variable as regular, one usually comes to infer a model belonging to the family of exponential distributions. This approach works fairly well to describe the central part of the income/wealth distribution, but does a poor job in describing the lower and upper ends. To explain the tails, principally the upper tail, the empirical distribution is often broken into two pieces by estimating the leftmost point (threshold) beyond which the distribution is better described by a parametric model belonging to the family of power-law distributions. Therefore, if one privileges the wholeness of the distribution he should accept a poor fit of the upper tail, whereas if he is interested in a better interpolation of such a tail he should accept a model composed of two distinct distributions. This is fully consistent with what observed above: income and wealth are not regular quantities; rather, they behave strangely and, as it will be discussed, $\kappa$-generalized models are the most comfortable with the description and interpretation of strange quantities because they unify two distinct statistical principles into a single probability distribution by activating one of the two depending on the orders of magnitude of the realizations.

The fundamental characteristic of strange quantities is \textit{a-symmetry},\footnote{The notion of ``\textit{a}-symmetry'' was developed by \cite{Landini2016} and it should not be misunderstood with that of ``asymmetry'', which is included in the former. Indeed, \textit{a}-symmetry means a lack of symmetry of the same statistical principle due to the coexistence of two distinct statistical principles over the whole support of the quantity: up to given orders of magnitude there is no \textit{exclusion principle}, while beyond such a limit an \textit{exclusion principle} exists. That is, according to a weak analogy with physics, it is like units following the Bose-Einstein statistics at low orders of magnitude (bosons) whereas at higher orders of magnitude they follow the Fermi-Dirac statistics (fermions).} suggesting their intrinsic \textit{relativism} according to a relativistic interpretation of statistical mechanics \cite{Kaniadakis2002,Kaniadakis2005,Kaniadakis2009a,Kaniadakis2009b,Kaniadakis2013}. Moreover, the determinants of such a \textit{strange} characteristic can be understood in terms of the fundamental categories of socioeconomic complex systems \cite{Landini2016}. 

Three main aspects can be isolated for \textit{a}-symmetry: structural, behavioral and formal. Structural \textit{a}-symmetry is due to the \textit{heterogeneity} of observation units populating systems at different levels of granularity. Behavioral \textit{a}-symmetry is due to \textit{direct interaction} among elementary units\textemdash i.e. in terms of transfers of portions of the quantity\textemdash and \textit{indirect interaction} among sub-systems\textemdash i.e. due to the transitions of observation units from a sub-system to another while bringing with themselves their own endowments. Formal \textit{a}-symmetry, in turn, is related to the asymmetric shape of the distribution as a representation of the system as a whole, an \textit{emergent phenomenon} due to the continuous interaction of heterogeneous constituents causing \textit{inequality} of the distribution. All such traits can be detected when looking at income and wealth as strange quantities.

While making inference on strange quantities, standard numeric transformations do not allow for the formalization of a unified model when two statistical principles (e.g. the exponential and the power-law families) dominate different parts of the support of a random variable. Therefore, as in relativity theory one can explain within a unified framework both the Galilean and the Einstein relativity using the Lorentz transform on the non-Euclidean Minkowski space-time, in much the same way the mathematics of deformations developed by \cite{Kaniadakis2002,Kaniadakis2005,Kaniadakis2009a,Kaniadakis2009b,Kaniadakis2013} allows for a unified explanation of both the regular and the strange quantities over their whole support.

Usually, if not aware that $X$ is a strange quantity one treats it \textit{regularly}, for instance by using (alone or in combination) standard numeric transformations like scaling on the average, max-min normalizations, logarithmic transformations or root filtering. The effect of such transformations is much less effective the more the quantity is strange. This is because they operate in the same way over the whole support of $X$ even though they should differently react to data with respect to their orders of magnitude, which hide the absence/presence of an exclusion principle.\footnote{To be true, the logarithmic transform usually performs well because it operates more strongly on very small/high values, but it is not enough to grasp the ``strangeness'' of strange quantities. On the other hand, it is not a coincidence that the $\kappa$-generalized model relies on the $\kappa$-logarithm and $\kappa$-exponential operators.} Therefore, even after such transformations, it often happens that the \textit{rarefied} tails of the inferred distribution are not well interpreted as the more \textit{dense} central part. The reason resides in the fact a \textit{regular} transformation\textemdash i.e. based on the \textit{regular mathematics}\textemdash cannot grasp the \textit{strangeness} of realizations obeying or not an exclusion principle that depends on their order of magnitude; to do this one needs a \textit{deformed mathematics}.

Models belonging to the $\kappa$-generalized family are the best performing ones because, by involving a deformed mathematics, they succeed where the others fail. Model \eqref{eq:Equation_2}, for instance, behaves \textit{classically} ($\kappa=0$) when the quantity is \textit{regular}, hence reconciling with the family of exponential distributions; in contrast, when the quantity is \textit{strange}, it behaves \textit{relativistically} ($0<\kappa<1$): as seen in Section \ref{sec:BasicProperties}, below a given order of magnitude the $\kappa$-generalized reconciles with a model from the exponential family of distributions, while above that it follows a Pareto power-law distribution. The interesting aspect is that \eqref{eq:Equation_2} is a single model, activating the exponential or power-law behavior depending on the order of magnitude of the underlying quantity realizations. This, we believe, makes $\kappa$-generalized models the most appropriate ones for inference on distributions of income and wealth as strange quantities.

% CONCLUDING REMARKS -------------------------------------------------------------------------------------

\section{Concluding remarks}
\label{sec:ConcludingRemarks}

This paper has provided a brief exposition of all that appeared in the recent literature about the $\kappa$-generalized distribution and its extensions, a new and fruitful set of statistical models for the size distribution of income and wealth developed by some of us over several years of collaborative and multidisciplinary research. Given that the distribution only began to appear in the econophysics literature in 2007, it is safe to predict that there will be many further applications. On the theoretical side, especially as regards economics, there are still some unresolved issues, including the definition of a theoretical model able to demonstrate the emergence of $\kappa$-generalized income/wealth distributions as the result of decentralized interactions of a large number of heterogeneous agents. Hence, the end of the story is yet to come.

% REFERENCES ---------------------------------------------------------------------------------------------

% EOF ----------------------------------------------------------------------------------------------------

\end{document}